\documentclass[%
 reprint,
 amsmath,amssymb,
 aps,
]{revtex4-2}

\usepackage{amsmath}
\usepackage{graphicx,epstopdf}
\usepackage{dcolumn}
\usepackage{bm}
\usepackage{xcolor}
\usepackage{float}
\usepackage{enumerate}
\usepackage{natbib}
\usepackage{multirow}

\begin{document}

\title{Effect of characteristic size on the collective phonon transport in crystalline GeTe}

\author{Kanka Ghosh}
\email{kanka.ghosh@u-bordeaux.fr}
\affiliation{University of Bordeaux, CNRS, Arts et Metiers Institute of Technology, Bordeaux INP, INRAE, I2M Bordeaux, 351 Cours de la libération, F-33400 Talence, France}
\author{Andrzej Kusiak}%
\affiliation{University of Bordeaux, CNRS, Arts et Metiers Institute of Technology, Bordeaux INP, INRAE, I2M Bordeaux, 351 Cours de la libération, F-33400 Talence, France}
 \author{Jean-Luc Battaglia}%
 \affiliation{University of Bordeaux, CNRS, Arts et Metiers Institute of Technology, Bordeaux INP, INRAE, I2M Bordeaux, 351 Cours de la libération, F-33400 Talence, France}


\begin{abstract}

We study the effect of characteristic size variation on the phonon thermal transport in crystalline GeTe for a wide range of temperatures using the first-principles density functional method coupled with the kinetic collective model approach. The characteristic size dependence of phonon thermal transport reveals an intriguing collective phonon transport regime, located in between the ballistic and the diffusive transport regimes. Therefore, systematic investigations have been carried out to describe the signatures of phonon hydrodynamics via the competitive effects between grain size and temperature. A characteristic non-local length, associated with phonon hydrodynamics and a heat wave propagation length has been extracted. The connections between phonon hydrodynamics and these length scales are discussed in terms of the Knudsen number. Further, the scaling relation of thermal conductivity as a function of characteristic size in the intermediate size range emerges as a crucial indicator of the strength of the hydrodynamic behavior. A ratio concerning normal and resistive scattering rates has been employed to understand these different scaling relations, which seems to control the strength and prominent visibility of the collective phonon transport in GeTe. This systematic investigation emphasizes the importance of the competitive effects between temperature and characteristic size on phonon hydrodynamics in GeTe, which can lead to a better understanding of the generic behavior and consequences of the phonon hydrodynamics and its controlling parameters in low-thermal conductivity materials.
\end{abstract}

\maketitle


\section{Introduction}

Detecting phonon hydrodynamics and associated collective phonon transport in low-thermal conductivity materials is a challenging task due to the not-so-overwhelming differences between the normal and the resistive phonon scattering rates. This leads to the exploration of very low cryogenic temperatures to see a visible effect of collective motion of phonons. On the other hand, 2D materials draw an appreciable amount of studies \cite{Cepellotti2015, Gill2015, Li_graphene2018} concerning phonon hydrodynamics because of their enhanced normal scattering phenomena. This helps in realizing phonon hydrodynamics even at higher temperatures and therefore can be understood using experiments. Nevertheless, collective phonon transport holds fundamental interest in materials as it draws parallel to the hydrodynamic flow in fluids. Investigating this collective phonon transport in low-thermal conductivity materials is crucial to understand the role of different competing effects that influences phonon hydrodynamics and invokes fundamental question on its generality and validity in both high and low conductivity materials. The complete understanding of the origin of this phenomena thus demands a systematic decoupling between various controlling parameters that dictate phonon hydrodynamics in materials.

Phonon hydrodynamics is a heat transport phenomena where the collective flow of phonons dominate the heat conduction in materials \cite{Leebookchapter1, Lucas2019, Cepellotti2015, Hardy, GUO20151, Machidaeaat3374}. This is enabled by significantly higher momentum conserving normal scattering (N) events compared to other dissipative scattering events [Umklapp (U), isotope (I) and boundary scattering (B)], favoring damped wave propagation of temperature fluctuations \cite{Bi2018, Transport_waves_as_crystal_excitations}. In their consecutive two pioneering theoretical works \cite{Guyer1966_1, Guyer1966_2} published in 1966, Guyer and Krumhansl distinguished the phonon hydrodynamics for nonmetallic crystals using the comparison between normal and resistive average scattering rates. Phonon hydrodynamics have also been realized by the deviation from Fourier's law at certain length and time scales \cite{Giorgia2014nanoletters, Gill2015}. The concept of the kinetic theory of relaxons to characterize phonon hydrodynamics have been introduced by Cepellotti and Marzari \cite{Relaxon}. Very recently, Sendra $\textit{et al.}$ \cite{sendra2021} introduced a framework to use hydrodynamic heat equations from phonon Boltzmann equation to study the hydrodynamic effects in semiconductors.     

Experiments and theoretical investigations over the years suggest that only few and mostly two-dimensional (2D) materials possess phonon hydrodynamics \cite{Lucas2019, Gill2015, Li_graphene2018, Li_graphene_2019, Derek_graphene_2018}. Some of these 2D materials like graphene and boron nitride \cite{Cepellotti2015} can even persist phonon hydrodynamics at room temperature due to the presence of strong normal scattering realized via first-principles simulations. Recently, the relation between the thickness and thermal conductivity and consequently their connection to the phonon hydrodynamics was studied for graphite \cite{Machida309}. The presence of second sound, a prominent manifestation of phonon hydrodynamics, was also observed in graphite at a temperature higher than 100 K via the experiments carried out by Huberman \textit{et al.} \cite{Huberman375}. This validates the predictions of the simulation studies done by Ding \textit{et al.} \cite{Ding2018} on graphite. Similarly, theoretical evaluations by Markov \textit{et al}. \cite{Bi2018} confirmed the experimental observation \cite{Bi1972} of hydrodynamic Poiseuille phonon flow in bismuth (Bi) at low temperature. A faster than $T^3$ scaling of the lattice thermal conductivity was described as a marker to identify phonon hydrodynamics in bulk black phosphorus \cite{Machidaeaat3374} and SrTiO$_{3}$ \cite{Koreeda2007, Strontium2018}. Koreeda $\textit{et al.}$ \cite{Koreeda2010} studied collective phonon transport in KTaO$_3$ using low frequency light-scattering and time-domain light-scattering techniques and phonon hydrodynamics was found to exist below 30 K . Further second sound was also observed in solid helium (0.6 - 1 K) \cite{Solid_Helium}, NaF ($\sim$ 15 K) \cite{NaF} at low temperatures.

As discussed earlier, the studies of phonon hydrodynamics for low-thermal conductivity materials are substantially less compared to its high-thermal conductivity counterpart. However, a systematic decoupling of various controlling parameters can help manipulate phonon hydrodynamic behavior in the low-thermal conductivity materials. Torres \textit{et al.} \cite{Torres_2019} showed a strong phonon hydrodynamic behavior in low-lattice thermal conductivity ($\kappa_L$) materials such as single layer transition metal dichalcogenides (MoS$_2$, MoSe$_2$, WS$_2$ and WSe$_2$). In our earlier paper \cite{kanka2}, we investigated the low temperature thermal transport in crystalline GeTe, a chalcogenide-based material of diverse practical interests \cite{Levin2013, Campi2015}, which shows even lower lattice thermal conductivity compared to metal dichalcogenides and found that it exhibits phonon hydrodynamics. However we found that the presence of hydrodynamic phonon transport in crystalline GeTe is sensitive to the grain size and vacancies present in the material. Further, temperature was found to play an important role in favoring appreciable normal scattering events to enable collective phonon transport.

For low-thermal conductivity materials like GeTe, the characteristic size of the material and temperature are two crucial parameters that influence the existence of phonon hydrodynamics. Distinguishing the competing effects of these two factors is important for general understanding of collective phonon transport in GeTe. Therefore, in the current paper, we investigate the effects of characteristic size ($L$) on the collective thermal transport in low-thermal conductivity crystalline GeTe for temperatures ranging from 4 K to around 500 K. We use first-principles calculations with a kinetic collective model approach \cite{KCM-method2017} for this paper. We first identify the $L$-regimes corresponding to ballistic and complete diffusive regimes. Then we explore the regime of collective phonon transport that comprises both ballistic and diffusive phonons. Average scattering rates have been used to identify phonon hydrodynamic regimes both in terms of temperature and characteristic size. Further, temperature and $L$-regimes are quantified using the Knudsen number obtained using two different length scales concerning phonon hydrodynamics. The prominent signature of phonon hydrodynamics in GeTe is found to depend on the scaling exponent of thermal conductivity as a function of $L$ in the intermediate $L$-regime where phonon transport shifts from ballistic to complete diffusive. The ratio of normal to resistive scattering rates at this $L$-regime seems to dictate the strength of the hydrodynamic behavior.

\section{Computational Details}{\label{section:computational}}
First-principles density functional methods are employed to optimize the structural parameters of crystalline GeTe (space group $R3m$). The details of the parameters for GeTe can be found in our earlier paper \cite{kanka}. The phonon lifetime is calculated using PHONO3PY \cite{Togo} software package. The supercell approach with finite displacement of 0.03 \AA{} is employed to obtain the harmonic (second order) and the anharmonic (third order) force constants, given via
\begin{equation}
    \Phi_{\alpha \beta} (l\kappa, l'\kappa') = \frac{\partial^2 \Phi}{\partial u_{\alpha} (l\kappa)\partial u_{\beta} (l'\kappa')}
\end{equation}
and
\begin{equation}
    \Phi_{\alpha \beta \gamma} (l\kappa, l'\kappa', l''\kappa'') = \frac{\partial^3 \Phi}{\partial u_{\alpha} (l\kappa)\partial u_{\beta} (l'\kappa') \partial u_{\gamma} (l''\kappa'')}
\end{equation}
respectively. Density functional method is implemented with QUANTUM-ESPRESSO \cite{qe} to calculate the forces acting on atoms in supercells. The harmonic force constants are approximated as \cite{Togo} 
\begin{equation}
   \Phi_{\alpha \beta} (l\kappa, l'\kappa') \simeq - \frac{F_{\beta} [l'\kappa'; \textbf{u} (l \kappa)]}{u_{\alpha} (l \kappa)} 
\end{equation}
where \textbf{F}[$l'$$\kappa'$; \textbf{u}($l$$\kappa$)] is atomic force computed at \textbf{r}($l'$ $\kappa'$) with an atomic displacement \textbf{u}($l\kappa$) in a supercell. Similarly, third order force constants are calculated using\cite{Togo} 
\begin{equation}
   \Phi_{\alpha \beta \gamma} (l\kappa, l'\kappa', l''\kappa'') \simeq - \frac{F_{\gamma} [l''\kappa''; \textbf{u} (l \kappa), \textbf{u} (l' \kappa')]}{u_{\alpha} (l \kappa)u_{\beta} (l' \kappa')} 
\end{equation}
where \textbf{F}[$l''$$\kappa''$; \textbf{u}($l$$\kappa$), \textbf{u}($l'$ $\kappa'$)] is the atomic force computed at \textbf{r}($l''$ $\kappa''$) with a pair of atomic displacements \textbf{u}($l\kappa$) and \textbf{u}($l'\kappa'$) in a supercell. These two sets of linear equations are solved using the Moore-Penrose pseudoinverse as is implemented in PHONO3PY \cite{Togo}.

Using the 2$\times$2$\times$2 supercell and finite displacement method, we obtain 228 supercell configurations with different pairs of displaced atoms, for the calculations of the anharmonic force constants. A larger 3$\times$3$\times$3 supercell is employed for the harmonic force constants calculation. For force calculations, the reciprocal space is sampled with a 3$\times$3$\times$3 k-sampling Monkhorst-Pack (MP) mesh \cite{MP} shifted by a half-grid distances along all three directions from the $\Gamma$- point. For the density functional calculations, the Perdew-Burke-Ernzerhof (PBE) \cite{PBE} generalized gradient approximation (GGA) is used as the exchange-correlation functional. Due to its negligible effects on the vibrational features of GeTe, as mentioned in earlier studies \cite{Shaltaf2009,Campi2017}, the spin-orbit interaction has been ignored. Electron-ion interactions are represented by pseudopotentials using the framework of the projector-augmented-wave (PAW) method \cite{PAW}. The Kohn-Sham (KS) orbitals are expanded in a plane-wave (PW) basis with a kinetic cutoff of 60 Ry and a charge density cutoff of 240 Ry as specified by the pseudopotentials of Ge and Te. The total energy convergence threshold has been kept at 10$^{-10}$ a.u. for supercell calculations. The imaginary part of the self-energy has been calculated using the tetrahedron method from which phonon lifetimes are obtained.  

\section{Lattice dynamics and Kinetic Collective Model (KCM)}{\label{section:lattice dynamics}}

In the theory of lattice dynamics, the crystal potential is expanded with respect to atomic displacements and the third-order coefficients associated with anharmonicity are used to calculate the imaginary part of the self-energy \cite{Togo}. Generally, in a harmonic approximation, phonon lifetimes are infinite whereas, anharmonicity in a crystal yields a phonon self-energy  $\Delta\omega_{\lambda}$ + $i\Gamma_{\lambda}$. The phonon lifetime ($\tau_{ph-ph}$) has been computed from the imaginary part of the phonon self energy as $\tau_{\lambda}$ = $\frac{1}{2\Gamma_{\lambda}(\omega_{\lambda})}$ using PHONO3PY \cite{Togo, Mizokami} from the following equation
\begin{widetext}
\begin{equation}
    \Gamma_{\lambda}(\omega_{\lambda}) = \frac{18\pi}{\hbar^2}\sum_{\lambda'\lambda''}\Delta\left( \textbf{q}+\textbf{q}'+\textbf{q}'' \right)\mid \Phi_{-\lambda\lambda'\lambda''}\mid ^{2}\{(n_{\lambda'}+n_{\lambda''}+1)\delta(\omega-\omega_{\lambda'}-\omega_{\lambda''}) + (n_{\lambda'}-n_{\lambda''})[\delta(\omega+\omega_{\lambda'}-\omega_{\lambda''}) - \delta(\omega-\omega_{\lambda'}+\omega_{\lambda''})]\}
\end{equation}
\end{widetext}
where $n_\lambda$ = $\frac{1}{exp(\hbar\omega_{\lambda}/k_{B}T)-1}$ is the phonon occupation number at the equilibrium. $\Delta\left(\textbf{q}+\textbf{q}'+\textbf{q}'' \right)$ = 1 if $\textbf{q}+\textbf{q}'+\textbf{q}'' = \textbf{G}$, or 0 otherwise. Here \textbf{G} represents reciprocal lattice vector. Integration over \textbf{q}-point triplets for the calculation is made separately for normal (\textbf{G} = 0) and umklapp processes (\textbf{G} $\neq$ 0) and therefore phonon umklapp ($\tau_U$) and phonon normal lifetime ($\tau_N$) have been distinguished. Using second-order perturbation theory, the scattering of phonon modes by randomly distributed isotopes ($\tau_{I}^{-1}$) is given by Tamura \cite{Tamura} as
\begin{equation}
\resizebox{1.0\hsize}{!}{$
    \frac{1}{\tau_{\lambda}^{I}(\omega)} = \frac{\pi \omega_{\lambda}^{2}}{2N}\sum_{\lambda'} \delta\left(\omega - \omega'_{\lambda} \right) \sum_{k} g_{k}|\sum_{\alpha}\textbf{W}_{\alpha}\left(k,\lambda \right)\textbf{W}_{\alpha}^{*}\left(k,\lambda \right)| ^{2} 
$}
\end{equation}
where $g_k$ is the mass variance parameter, defined as 
\begin{equation}
    g_{k} = \sum_{i} f_{i} \left( 1 - \frac{m_{ik}}{\overline{m}_k}\right)^{2}
\end{equation}
$f_i$ is the mole fraction, $m_{ik}$ is the relative atomic mass of $i$th isotope, $\overline{m}_k$ is the average mass = $\sum_{i} f_{i} m_{ik}$, and $\textbf{W}$ is a polarization vector. The database of the natural abundance data for elements \cite{Laeter} is used for the mass variance parameters. The phonon-boundary scattering has been implemented using Casimir diffuse boundary scattering \cite{kaviany_2014} as $\tau_{\lambda}^{B}$ = $\frac{L}{\mid \textbf{v}_{\lambda} \mid}$, where, $\textbf{v}_{\lambda}$ is the average phonon group velocity of phonon mode $\lambda$ and $L$ is the grain size, which is also called Casimir length, the length phonons travel before the boundary absorption or re-emission \cite{kaviany_2014}.

We use the kinetic collective model (KCM) \cite{KCM-method2017} to obtain the lattice thermal conductivity of GeTe. The KCM method has emerged as a useful approach to depict heat transport at all length scales with the computational cost being substantially less than that of the full solution of the linearized Boltzmann transport equation. According to the KCM method, the heat transfer process occurs via both collective phonon modes, emerges from the normal scattering events and via independent phonon collisions. Therefore, lattice thermal conductivity can be expressed as a sum of both kinetic and collective contributions weighed by a switching factor ($\Sigma \in \left[0,1\right]$), which indicates the relative weight of normal and resistive scattering processes \cite{KCM-method2017, Torres_2019}. While each mode exhibits individual phonon relaxation time in the kinetic contribution, the collective contribution is designated by an identical relaxation time for all modes \cite{alvarez2018thermalbook, KCM-method2017}. In the kinetic contribution term, the boundary scattering is included via the Matthiessen's rule as \begin{equation}
    \tau_{k}^{-1} = \tau_{U}^{-1} + \tau_{I}^{-1} + \tau_{B}^{-1}
\end{equation}
where $\tau_{k}$ is the total kinetic phonon relaxation time. On the contrary, a form factor $F$, calculated from the sample geometry, is employed to incorporate boundary scattering in the collective term \cite{KCM-method2017, alvarez2018thermalbook}. The KCM equations are:
\begin{equation}
    \kappa_L = \kappa_{k} + \kappa_{c}
\end{equation}
\begin{equation}
    \kappa_k = (1-\Sigma)\int \hbar \omega \frac{\partial f}{\partial T}v^{2} \tau_{k} D \textit{d}\omega 
\end{equation}
\begin{equation}
    \kappa_c = (\Sigma F)\int \hbar \omega \frac{\partial f}{\partial T}v^{2} \tau_{c} D \textit{d}\omega 
\end{equation}

\begin{equation}
    \Sigma = \frac{1}{1+\frac{\langle\tau_{N}\rangle}{\langle\tau_{RB}\rangle}}
\end{equation}
 
where $\kappa_k$ and $\kappa_c$ are kinetic and collective contributions to $\kappa_L$, respectively. $\langle\tau_{N}\rangle$ and $\langle\tau_{RB}\rangle$ designate average normal phonon lifetime and average resistive (considering $U$, $I$, and $B$) phonon lifetimes, respectively. $\langle \tau_{N} \rangle$ and $\langle \tau_{RB}\rangle$ are defined in the KCM \cite{KCM-method2017} as integrated mean-free times,
\begin{equation}
    \langle \tau_{RB} \rangle = \frac{\int C_{1} \tau_k d\omega}{\int C_{1} d\omega}
\end{equation}
and 
\begin{equation}
    \langle \tau_{N} \rangle = \frac{\int C_{0} \tau_N d\omega}{\int C_{0} d\omega}
\end{equation}
where $\tau_k$ is the total kinetic relaxation time and phonon distribution function in the momentum space, represented in terms of $C_{i = 0, 1}(\omega)$, defined in Ref. \cite{KCM-method2017} as
\begin{equation}
    C_{i}(\omega) = \left( \frac{v |q|}{\omega}\right)^{2i} \hbar \omega \frac{\partial f}{\partial T} D
\end{equation}
where $v(\omega)$ is the phonon mode velocity and $\mid q\mid$ is modulus wave vector. $C_{0}$ represents the specific heat of mode $\omega$. $f$ stands for Bose-Einstein distribution function, $v$ is mode velocity and $D(\omega)$ is phonon density of states for each mode. $\tau_{c}$ denotes the total collective phonon relaxation time and defined as \begin{equation}
    \tau_{c} (T) = \frac{\int C_{1}d\omega}{\int (\tau_{I}^{-1}+\tau_{U}^{-1})C_{1}d\omega}
\end{equation}
$\Sigma$ stands for the switching factor. $F$ is the form factor approximated via \cite{alvarez2018thermalbook}
\begin{equation}
F(L_{eff}) = \frac{L_{eff}^2}{2\pi^2l^{2}} \left(\sqrt{1+\frac{4\pi^{2}l^2}{L_{eff}^2}}-1\right)   
\end{equation}
where, $L_{eff}$ is the effective length of the sample (in our system, we use $L_{eff}$ = $L$, the grain size) and $l$ is the characteristic non-local scale \cite{Guyer1966_1, alvarez2018thermalbook}. This characteristic non-local length $l$ emerges from the complete hydrodynamic description of the KCM and is defined as a parameter that determines the non-local range in phonon transport. In our earlier paper \cite{kanka2}, comparing the results for thermal conductivity obtained using both direct solutions of linearized Boltzmann transport equation (LBTE) and KCM for GeTe, we found an excellent agreement between them at low temperature. At higher temperatures, a reasonable matching trend is retrieved, with KCM exhibiting slightly lower values than the LBTE solutions. However, in the low temperature hydrodynamic regime for GeTe, the solutions of LBTE and KCM collapse satisfactorily. For all KCM \cite{KCM-method2017} calculations of lattice thermal conductivity and associated parameters, KCM.PY code \cite{KCM-method2017} is implemented with the outputs obtained using PHONO3PY \cite{Togo}.

\section{Results and Discussions}{\label{section:Results and Discussions}}

\subsection{Ballistic and diffusive phonon transport}
\begin{figure}[H]
\centering
\includegraphics[width=0.5\textwidth]{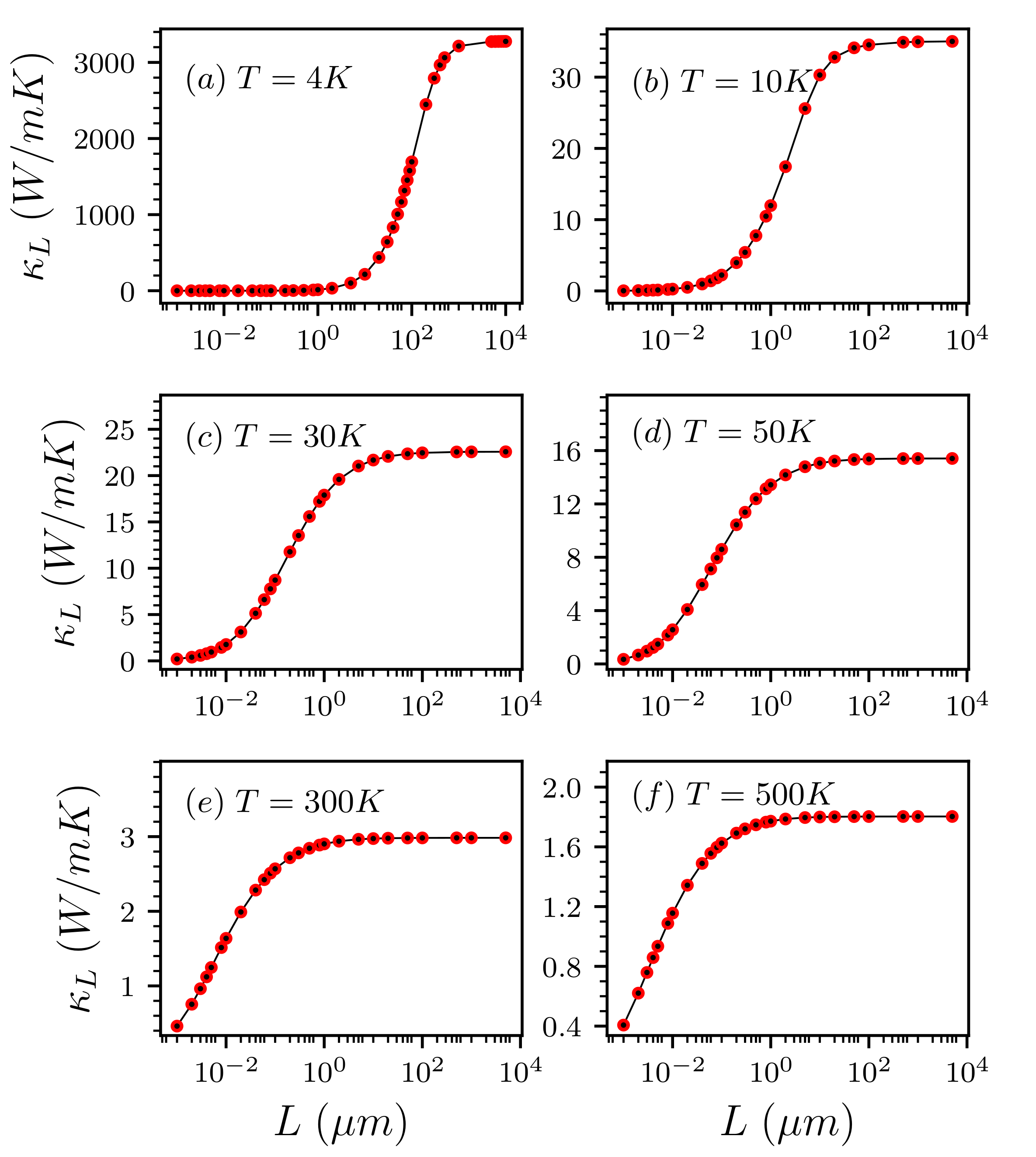}
\caption{The variation of lattice thermal conductivity ($\kappa_L$) as a function of characteristic length ($L$) of the GeTe sample at different temperatures.}
\label{fig:kappa_L_variation}
\end{figure}

As a first step to elucidating the complex collective behavior of phonons as a function of characteristic size ($L$), it is imperative to explore the variation of $\kappa_L$ with $L$ and therefore to identify the effect of $L$ on the ballistic and diffusive phonon transport. Figure \ref{fig:kappa_L_variation} describes this   

\noindent variation of GeTe for a wide temperature range (4 - 500 K). As the $L$ varies almost $10^6$ orders of magnitude (from 0.001 $\mu$m to 5000 $\mu$m), $\kappa_L$ undergoes a transition from a linear variation of $L$ to a plateau-like regime, and corresponds to complete ballistic and complete diffusive transport respectively. As we gradually go from lower to higher temperatures, the ballistic regime shrinks and the diffusive regime starts growing. Also, the onset of diffusive transport gradually seems to take place at lower 
\begin{figure}[H]
    \centering
    \includegraphics[width=0.5\textwidth]{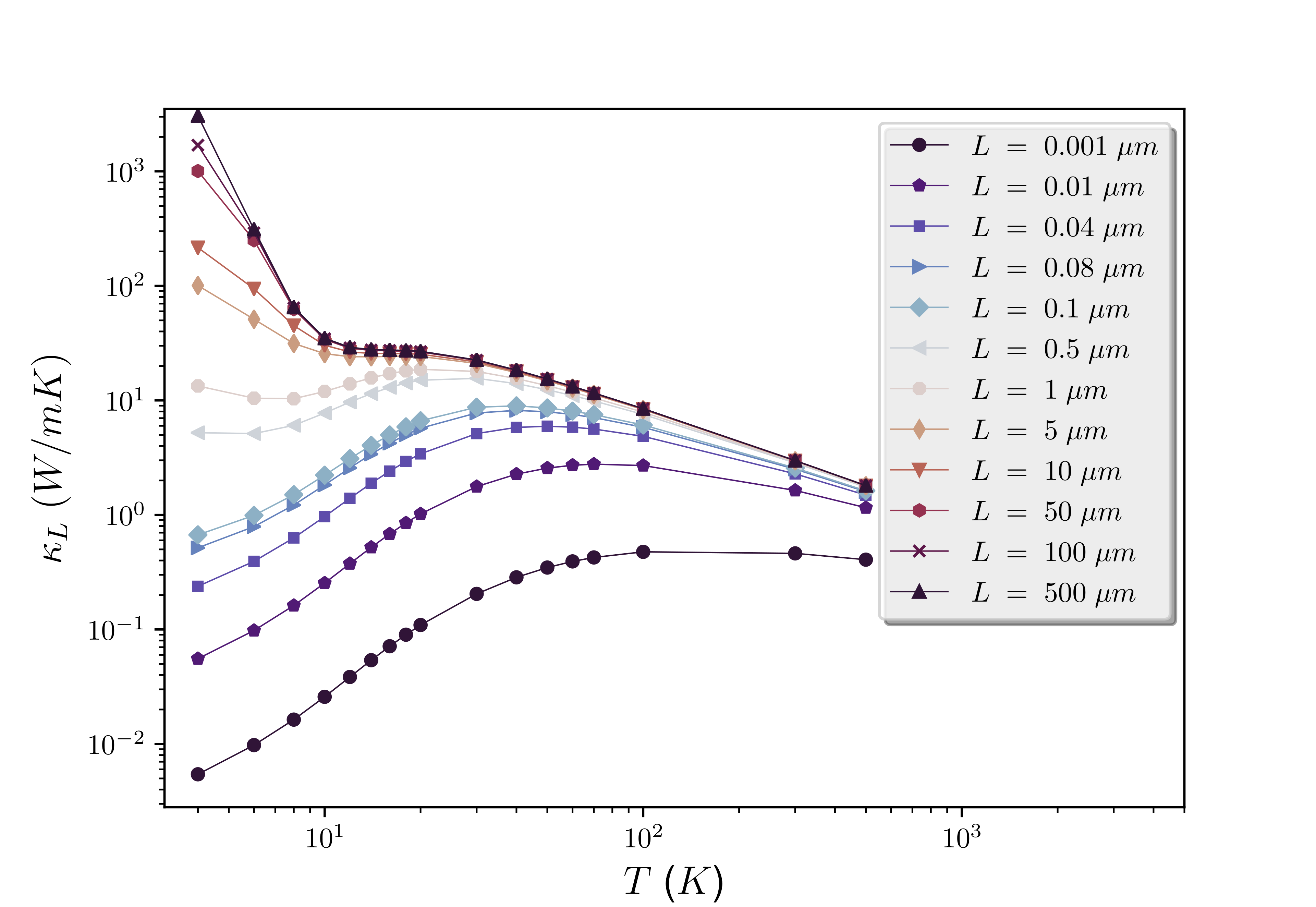}
    \caption{The variation of lattice thermal conductivity ($\kappa_L$) with temperature as a function of $L$ for crystalline GeTe.}
    \label{fig:kappa}
\end{figure}

\onecolumngrid
\begin{widetext}
\begin{figure}[H]
    \centering
    \includegraphics[width=1.0\textwidth]{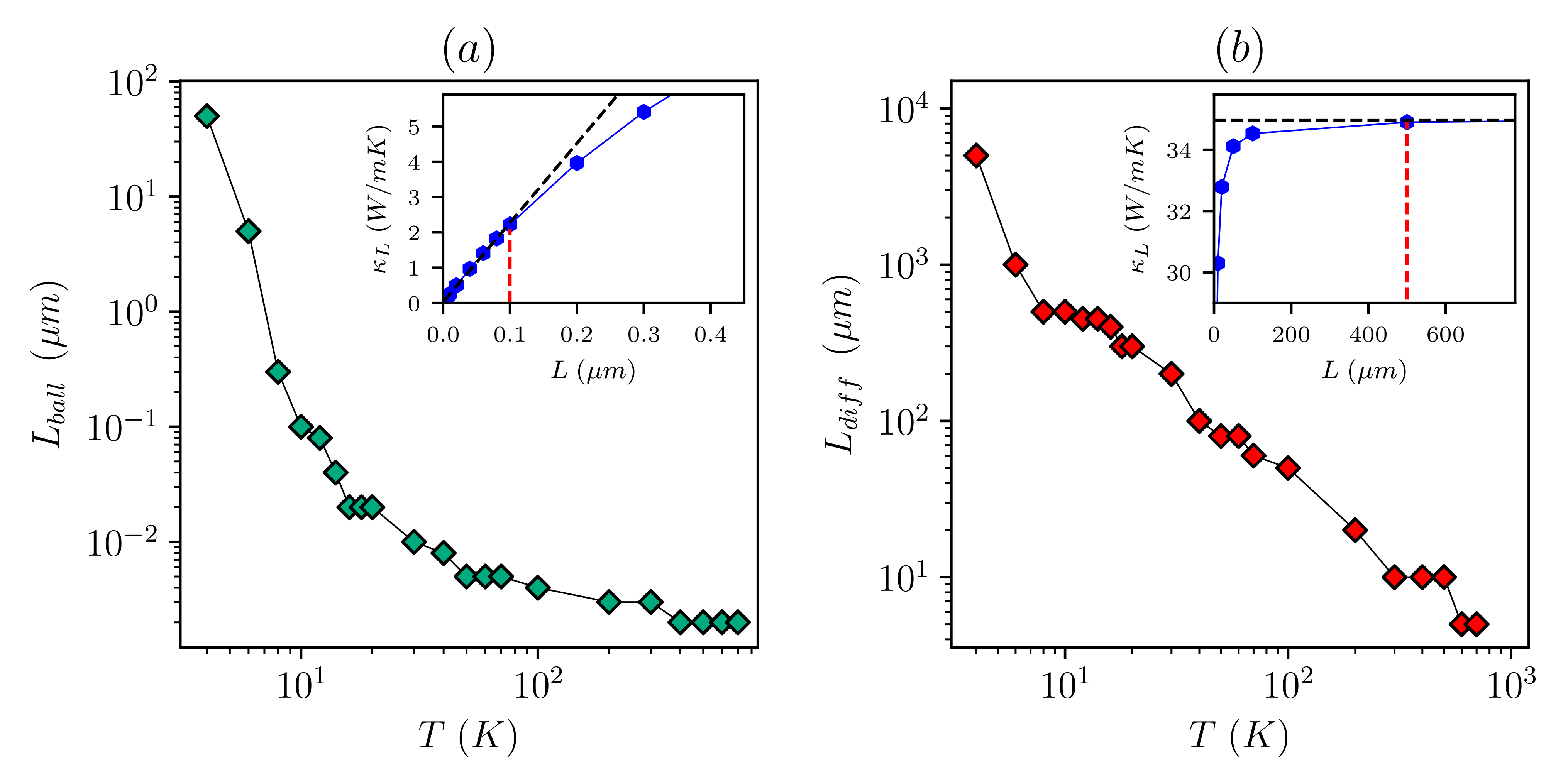}
    \caption{The variation of (a) $L_{ball}$ and (b) $L_{diff}$ are represented as a function of temperature for crystalline GeTe. The insets of (a) and (b) display the defining procedure of $L_{ball}$ and $L_{diff}$ respectively for a representative case of $T$ = 10 K.}
    \label{fig:L_ball_L_diff}
\end{figure}
\vspace{-0.3cm}
\end{widetext}
values of $L$ as we increase the temperature. It is well known in the literature \cite{Giorgia2014nanoletters, PhysRevApplied.9.024017} that ballistic conduction of phonons occurs without ph-ph scattering and displays a linear variation with $L$, whereas diffusive conduction of phonons manifests when scattered phonons carry the heat. The effect of the characteristic size on $\kappa_L$ can also be represented via the variation of $\kappa_L$ with temperatures for different $L$, as shown in Fig 2. At higher temperatures, it is well known \cite{kanka} that $\kappa_L$ decreases with $T$, with 1/$T$ scaling due to the dominant umklapp scattering at high temperatures. As temperature is lowered, gradually $\kappa_L$ attains a peak following a gradual decrement at very low temperature. As we go towards higher $L$, the peaks of $\kappa_L$ as a function of temperature are gradually seen to be shifted towards lower temperatures (Fig. \ref{fig:kappa}).

\noindent The effect of $L$ on the temperature variation of $\kappa_L$ gives rise to an interesting feature as we increase $L$ above a certain limit. It is known that $L$ plays a crucial role via phonon-boundary scattering as gradual increment of $L$ assists in weakening the boundary scattering. This weakening of boundary scattering and strong normal scattering rates (to be discussed later) at low temperatures transforms the peak of $\kappa_L$ into a cusp-like feature when $L$ $\geq$ 1 $\mu$m and $\kappa_L$ is further seen to be increased at very low temperatures.

To give a more precise account of ballistic and diffusive conduction of phonons in GeTe, we further investigate the characteristic size range of ballistic and diffusive conduction as a function of temperature. The complete ballistic length regime ($L_{ball}$) is defined via the maximum value of $L$, until which $\kappa_L$ varies linearly with $L$. Similarly, the complete diffusive length regime ($L_{diff}$) is defined via the minimum length $L$, above which $\kappa_L$ reaches the thermodynamic limit and therefore reaches a plateau. In other words, $L_{diff}$ represents the longest mean free path of the heat carriers at a particular temperature \cite{Giorgia2014nanoletters}. Figures \ref{fig:L_ball_L_diff}.(a) and \ref{fig:L_ball_L_diff}.(b) represent the variations of $L_{ball}$ and $L_{diff}$ respectively, as a function of temperature. As temperature increases, we see a gradual decrement of both $L_{ball}$ and $L_{diff}$. We note here that at very high temperatures, we hardly observe any ballistic conduction of phonons and the $L_{diff}$ acquires a very low value. This is representative of the fact that at high temperatures, internal phonon-phonon scattering is so dominant that no ballistic heat conduction is seen to exist, even for very small grains of the order of 1 nm.

To delve deeper into the origin of length dependent $\kappa_L$ in the ballistic phonon conduction regime of GeTe, we investigate the contribution of acoustic and optical modes in the ballistic propagation of heat. Earlier, molecular dynamics simulations and experiments on suspended single-layer graphene \cite{doi:10.1063/1.4817175, Xu2014} suggested the ballistic propagation of long-wavelength, low-frequency acoustic phonon to be solely responsible for the length-dependent $\kappa_L$ in the ballistic regime. Our previous studies on GeTe \cite{kanka, kanka2} suggested that GeTe shows a clear distinction between acoustic and optical modes in the frequency domain around 2.87 THz. The density of states goes to zero around a frequency of 2.87 THz \cite{kanka}, distinguishing two distinct frequency regimes: acoustic regime ($\omega$ $<$ 2.87 THz) and optical regime ($\omega$ $>$ 2.87 THz). We calculate the cumulative lattice thermal  

\onecolumngrid
\begin{widetext}
\begin{figure}[H]
    \centering
    \includegraphics[width=1.0\textwidth]{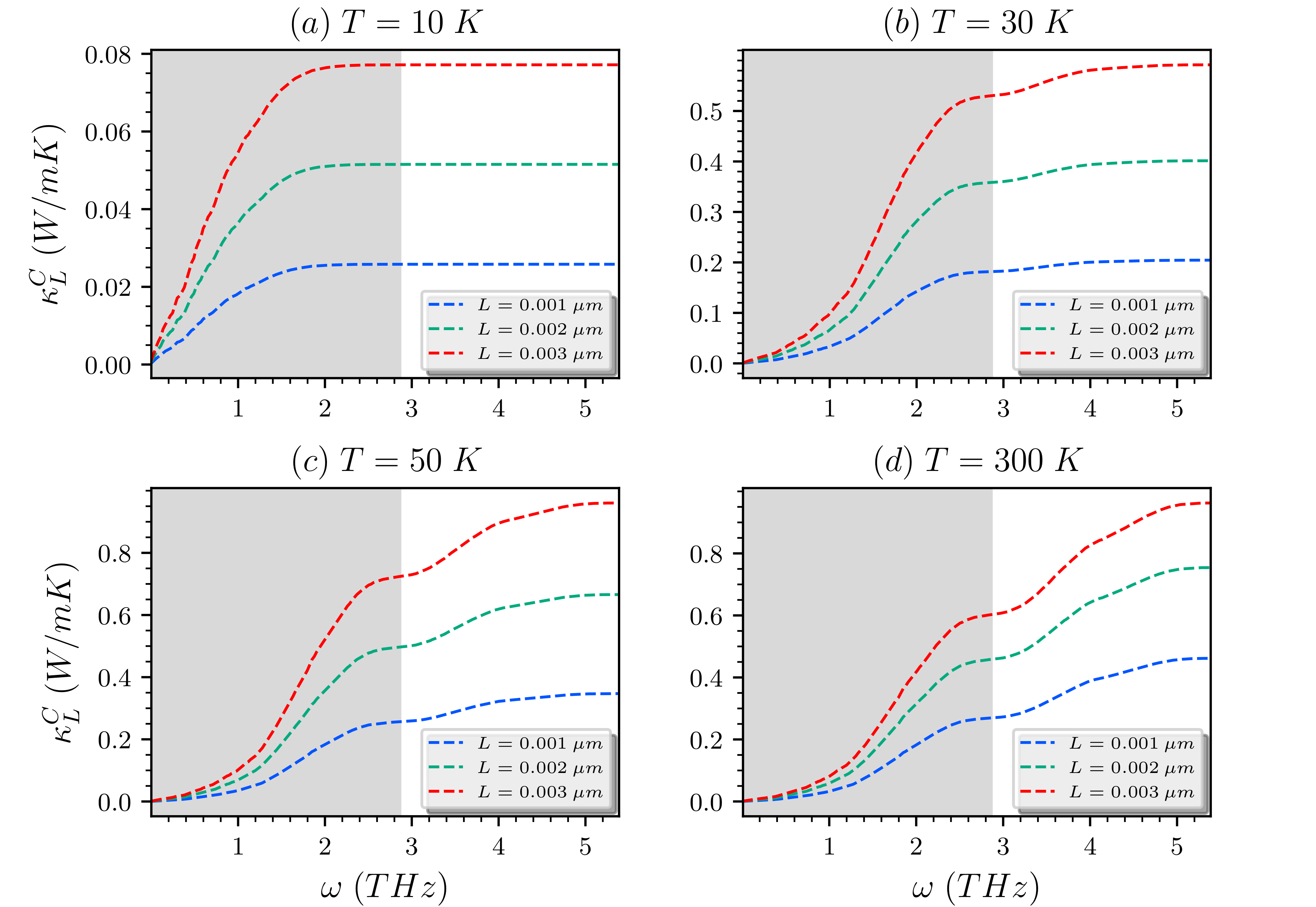}
    \caption{The variation of cumulative lattice thermal conductivity ($\kappa_{L}^{c}$) of crystalline GeTe as a function of phonon frequency ($\omega$) for four different temperatures: (a) 10 K, (b) 30 K, (c) 50 K and (d) 300 K. For each temperature, the variation of $\kappa_{L}^{C}$ with $\omega$ is presented for three different $L$: 0.001, 0.002 and 0.003 $\mu$m. The gray shaded region denotes the acoustic modes regime for GeTe.}
    \label{fig:AM_OM_contrib_ballistic}
\end{figure}
\vspace{-0.5cm}
\end{widetext}

\noindent conductivity ($\kappa_{L}^{c}$) as a function of phonon frequency defined as \cite{Togo,Mizokami}
\begin{equation}
    \kappa_L^c = \int_0^{\omega} \kappa_{L} (\omega')d\omega'
\end{equation}
where $\kappa_{L}$ ($\omega'$) is defined as \cite{Togo,Mizokami}
\begin{equation}
    \kappa_{L}(\omega') \equiv \frac{1}{NV_0} \sum_{\lambda} C_{\lambda} \textbf{v}_{\lambda} \otimes \textbf{v}_{\lambda}\tau_{\lambda} \delta (\omega'-\omega_{\lambda}) 
\end{equation}
with $\frac{1}{N}$ $\sum_{\lambda} \delta(\omega' - \omega_{\lambda})$ the weighted density of states (DOS). Figure \ref{fig:AM_OM_contrib_ballistic} presents the variation of average $\kappa_{L}^{c}$ (= $\frac{(\kappa_{xx}^{C}+\kappa_{yy}^{C}+\kappa_{zz}^{C})}{3}$) with phonon frequency. The density of states goes to zero at a frequency where $\kappa_L^c$ reaches a plateau defining the separation between acoustic (frequency $<$ 2.87 THz) and optical (frequency $>$ 2.87 THz) modes. Except at low temperature ($T$ = 10 K), the contribution from optical modes seem to present at all other temperatures. As we gradually increase the temperature [from Fig \ref{fig:AM_OM_contrib_ballistic}.(b) to Fig \ref{fig:AM_OM_contrib_ballistic}.(d)], the contributions from optical modes are seen to be enhanced. For example, for $L$ = 0.003 $\mu$m, the contribution of optical modes at $T$ = 10 K, 30 K, 50 K and 300 K are 0 $\%$, 9.9 $\%$, 24.2 $\%$, and 37.7 $\%$ respectively. Therefore, contrary to the understanding of ballistic propagation for a 2D material like single-layer graphene, except for very low temperatures, GeTe also shows a weak contribution from optical modes in the ballistic phonon propagation regime. However, the significant contributions come from acoustic modes in this regime.

\begin{figure}[H]
    \centering
    \includegraphics[width=0.5\textwidth]{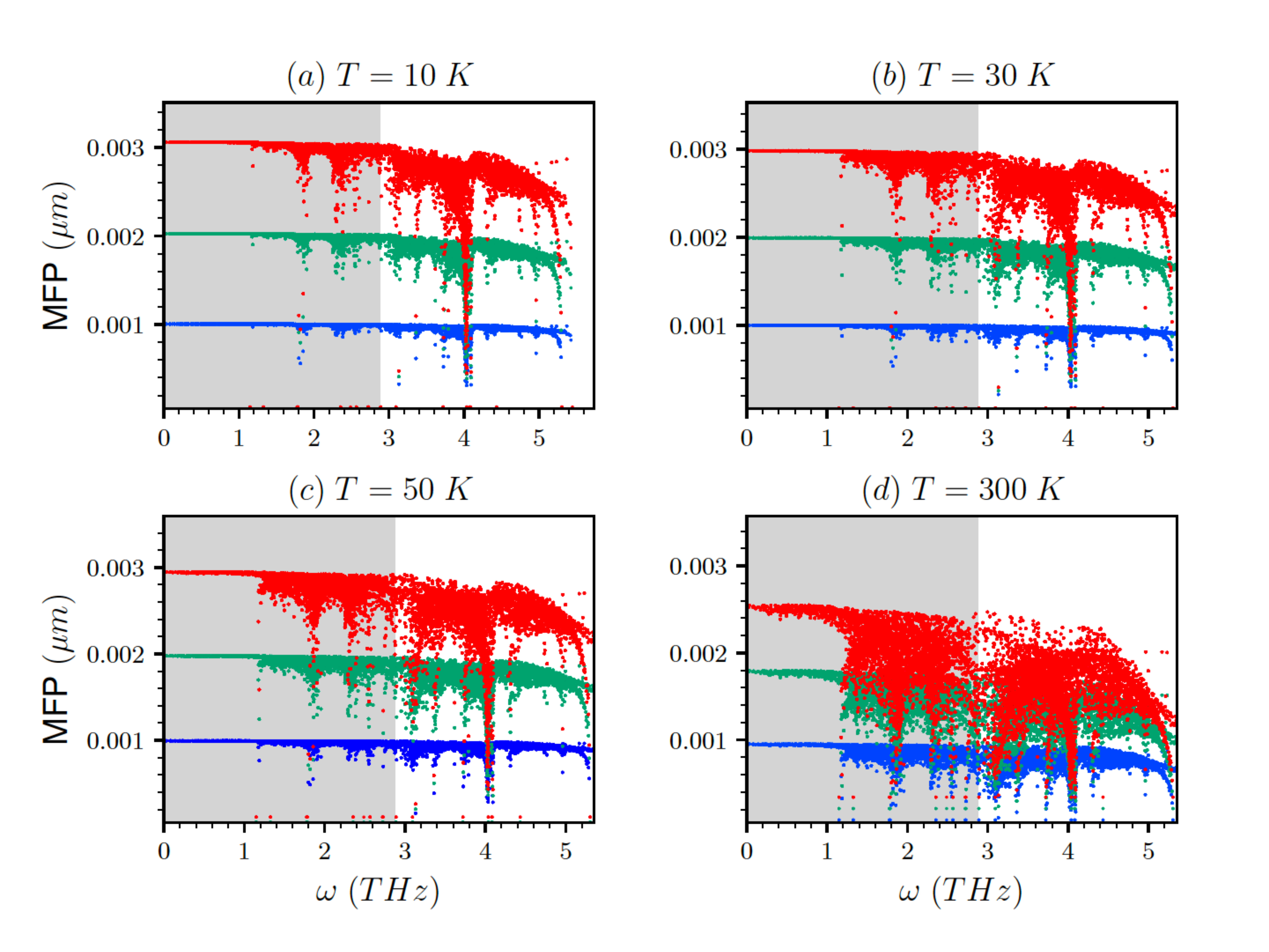}
    \caption{Effective mean free path (MFP) of crystalline GeTe are presented as a function of frequencies for three different $L$: 0.001 $\mu$m (blue points), 0.002 $\mu$m (green points) and 0.003 $\mu$m (red points) at four different temperatures: (a) $T$ = 10 K, (b) $T$ = 30 K, (c) $T$ = 50 K, (d) and $T$ = 300 K. The gray shaded region denotes the acoustic modes regime for GeTe.}
    \label{fig:mfp}
\end{figure}

To visualize the consequences on the mean-free path of the phonons at small $L$, we present the variation of the effective mean-free path variation with phonon frequency for different $L$ at different temperatures in Fig. \ref{fig:mfp}. In the KCM nomenclature, the kinetic mean free path [$l_{k}(\omega)$] and the collective mean free path [$l_{c}(T)$] are defined as $l_{k}(\omega)$ = $v\tau_k$ and $l_{c}(T)$ = $\overline{v}\tau_c$ respectively, where $v$ is the group velocity and 
\begin{equation}
    \overline{v} = \frac{\int v\hbar\omega\frac{\partial f}{\partial T}D(\omega)d\omega}{\int \hbar\omega\frac{\partial f}{\partial T}D(\omega)d\omega}
\end{equation}
is the mean phonon velocity \cite{KCM-method2017}. As the kinetic MFP 

is a function of phonon frequency whereas the collective MFP is frequency independent and varies only with temperature, we present an effective MFP as $l_{eff} (\omega)$ = (1-$\Sigma$)$l_{k}(\omega)$+$\Sigma$ $l_{c}$. The separate contributions from collective and kinetic MFPs are described in Supplemental Fig. S1. Two effects can be observed from this representation. First, at low temperature, as the $L$ is increased, the optical modes at higher frequencies exhibit more scattered mean-free paths. Figure \ref{fig:mfp}.(a) shows that at $T$ = 10 K, at higher frequencies in the optical modes, $L$ = 0.003 $\mu$m persists more scattered MFPs compared to the $L$ = 0.001 $\mu$m case. This feature indicates that the ballistic conduction is stronger for $L$ = 0.001 $\mu$m, where $L$ strongly controls the mean-free path than that of the $L$ = 0.003 $\mu$m case. Second, increasing temperature for fixed $L$, also leads to the gradual weakening of the ballistic conduction of phonons, as can be seen from Fig. \ref{fig:mfp}. This is evident from the gradual broadening of MFPs with temperature [follow fixed color points for four different temperatures in Fig. \ref{fig:mfp}.(a), (b), (c) and (d).] due to the gradually weakening control of $L$ on dictating the mean-free paths of the system.

\subsection{Collective phonon transport}

After understanding the effect of characteristic size ($L$) on the ballistic and diffusive conduction of phonons, we turn our attention to the effect of $L$ on the collective phonon transport of crystalline GeTe. The connection between ballistic and diffusive phonon transport and the collective motion of phonons are crucial to determine the origin of the exotic hydrodynamic phonon transport in materials. In our earlier work \cite{kanka2}, unusually, low-thermal conductivity chalcogenide GeTe emerged as a possible candidate to feature phonon hydrodynamics with the characteristic size being a dominant factor.
 
\onecolumngrid
\begin{widetext}
\vspace*{-1cm}
\begin{figure}[H]
    \centering
    \includegraphics[width=1.0\textwidth]{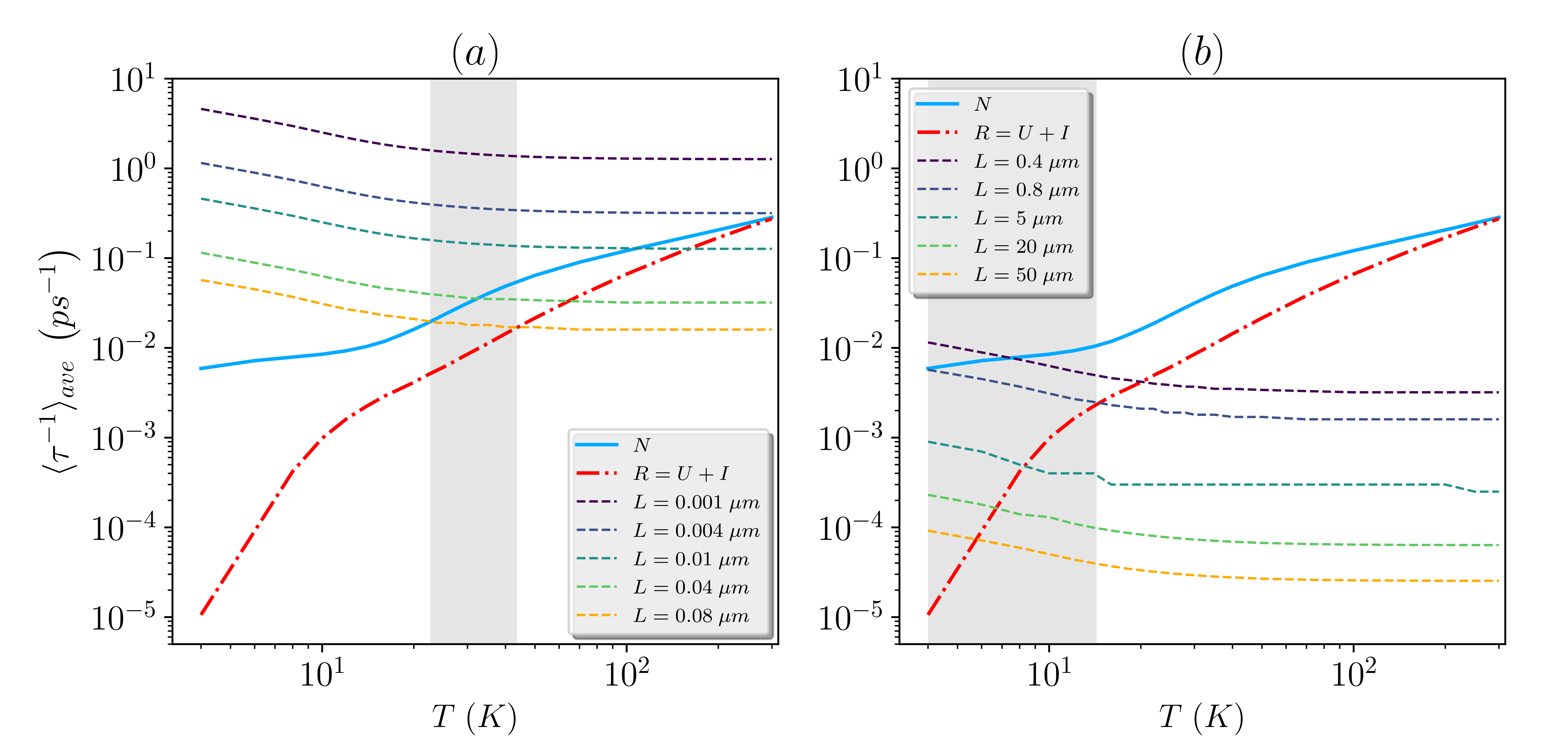}
    \caption{Thermodynamic average phonon scattering rates as a function of temperature for GeTe for different characteristic sizes ($L$). $N$, $U$, $I$ and $R$ denote normal, umklapp, isotope and resistive scattering respectively. Boundary scattering rates for different $L$ are also presented. The shaded regions in (a) and (b) correspond to the validation of the Guyer's condition \cite{Guyer1966_2} for Poiseuille's flow (Eq. \ref{eq:poiseuille}) for $L$ = 0.08 $\mu$m and $L$ = 0.8 $\mu$m respectively.}
    \label{fig:scattering_rate}
\end{figure}
\vspace{-0.5cm}
\end{widetext}

\onecolumngrid
\begin{widetext}
\begin{figure}[H]
    \centering
    \includegraphics[width=1.0\textwidth]{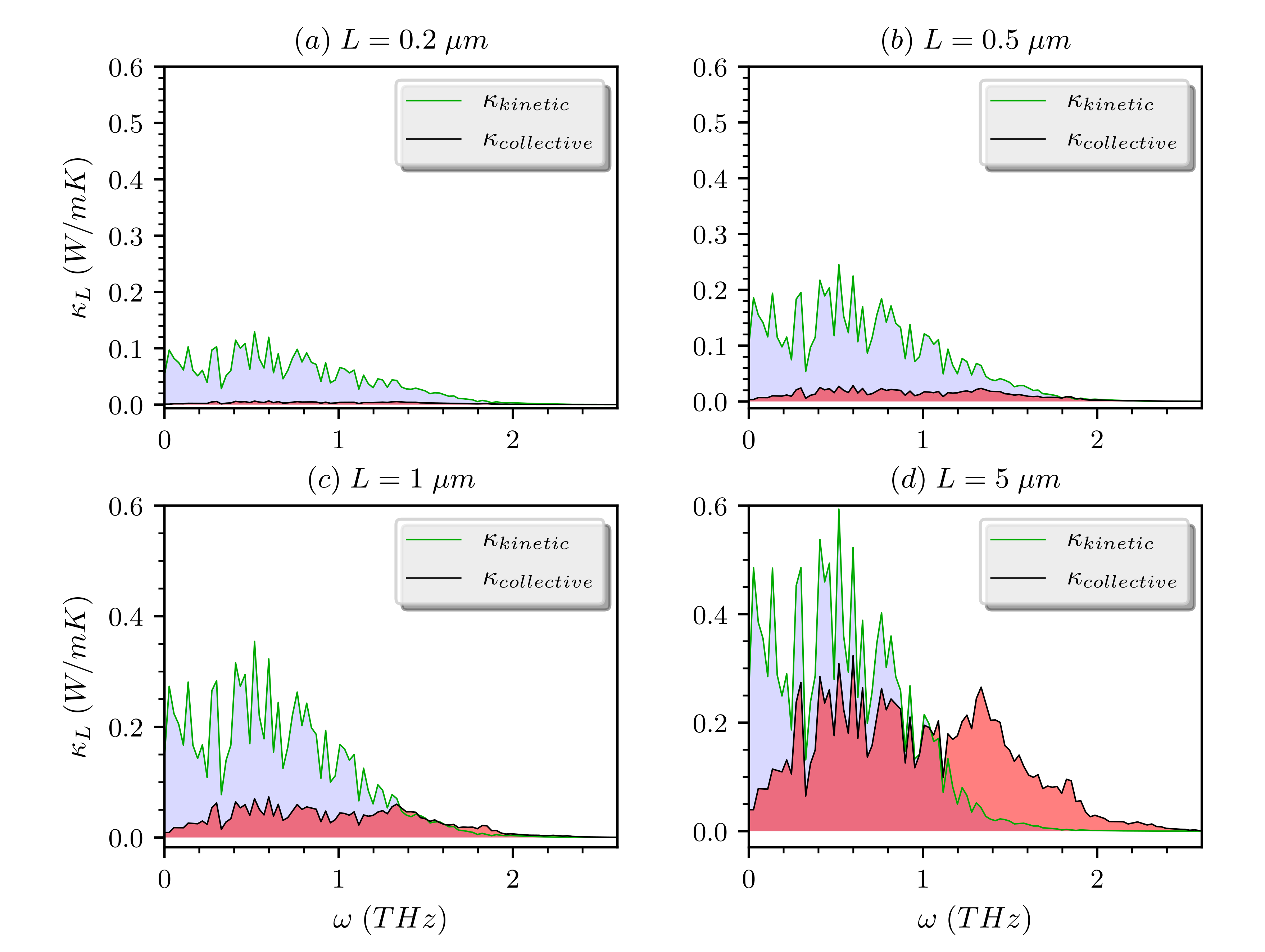}
    \caption{The spectral representation of lattice thermal conductivity ($\kappa_L$) as a function of phonon frequency at $T$ = 10 K for four different characteristic size or grain sizes ($L$): (a) 0.2 $\mu$m, (b) 0.5 $\mu$m, (c) 1 $\mu$m and (d) 5 $\mu$m. The kinetic contribution ($\kappa_{kinetic}$) is defined using light violet and the collective contribution ($\kappa_{collective}$) is defined using light brown color.}
    \label{fig:kl_collective}
\end{figure}
\vspace{-0.5cm}
\end{widetext} 
 
In this context, we start by investigating the relative strengths of the average phonon scattering rates, which is defined as
\begin{equation}
    \langle \tau_{i}^{-1} \rangle_{ave} = \frac{\sum_{\lambda}C_{\lambda}\tau_{\lambda i}^{-1}}{\sum_{\lambda}C_{\lambda}}
\end{equation}
Here, $\lambda$ denotes phonon modes ($\textbf{q}$, $j$) comprising wave vector $\textbf{q}$ and branch $j$. Index $i$ denotes normal, umklapp, isotope, and boundary scattering processes used, marked by N, U, I, and B respectively. $C_\lambda$ is the modal heat capacity, given by the following equation 
\begin{equation}
    C_\lambda = k_{B} \left(\frac{\hbar\omega_{\lambda}}{k_{B}T}\right)^{2} \frac{exp(\hbar\omega_{\lambda}/k_{B}T)}{[exp(\hbar\omega_{\lambda}/k_{B}T) -1]^2}
\end{equation}
where, $T$ denotes temperature, $\hbar$ is the reduced Planck's constant, and $k_B$ is the Boltzmann constant. In one of the earliest works on phonon hydrodynamics, Guyer and Krumhansl \cite{Guyer1966_2} found that the hydrodynamic regime exists if
\begin{equation}{\label{eq:hydro}}
   \langle \tau_{U}^{-1} \rangle_{ave} \ll \langle \tau_{N}^{-1} \rangle_{ave}
\end{equation}
Further, Guyer's condition \cite{Guyer1966_2} for the presence of second sound and Poiseuille's flow reads: 
\begin{equation}{\label{eq:poiseuille}}
   \langle \tau_{U}^{-1} \rangle_{ave} < \langle \tau_{B}^{-1} \rangle_{ave} < \langle\tau_{N}^{-1} \rangle_{ave}
\end{equation}

In Fig \ref{fig:scattering_rate}, we explore the $L$ window that satisfies the aforementioned Guyer and Krumhansl condition of phonon hydrodynamics in crystalline GeTe. Figure \ref{fig:scattering_rate} presents the average scattering rates due to normal (N), resistive (R) [comprised of umklapp (U) and isotope scattering (I)] and the phonon-boundary scattering as a function of temperature for GeTe. We observe a substantial width of $L$, that persists phonon hydrodynamic conditions, as the boundary scattering rates decrease gradually on increasing $L$. This is shown via the gray shaded regions in Figs. \ref{fig:scattering_rate}.(a) and \ref{fig:scattering_rate}.(b) for two representative grain sizes: $L$ = 0.08 $\mu$m and $L$ = 0.8 $\mu$m, respectively. In the scattering rate approach, we also identified the ballistic conduction region, mentioned earlier through the linear dependence of $\kappa_L$ with $L$, as the region where  $\langle \tau_{B}^{-1} \rangle_{ave} \gg \langle \tau_{ph-ph}^{-1} \rangle_{ave}$. Similarly, the purely diffusive conduction region, mentioned earlier as the $L$-regime where $\kappa_L$ is independent of $L$, as the region where $\langle \tau_{B}^{-1} \rangle_{ave} \ll \langle \tau_{ph-ph}^{-1} \rangle_{ave}$. At this point, we go back to Fig. \ref{fig:kappa} to explain the cusp-like behavior of $\kappa_L$ as a function of temperature. This cusp-like pattern of $\kappa_L$ is found to present for $L$ $>$ 1 $\mu$m, as we gradually decrease the temperature. In Fig. \ref{fig:scattering_rate}.(b), this $L$ regime is identified as $L$ values for which normal scattering overpowers boundary scattering rates. At low temperatures, umklapp scattering is rare and boundary scattering acts as the dominant resistive phonon scattering. So, the effect of boundary scattering tries to reduce the $\kappa_L$ while the momentum conserving normal scattering tries to increase $\kappa_L$. Overpowering normal scattering compared to boundary scattering for $L$ $>$ 1 $\mu$m forces $\kappa_L$ to increase after an apparent shallow dip or a plateau and gives rise to the cusp-like pattern in Fig. \ref{fig:kappa}.

Once the Guyer and Krumhansl conditions are satisfied and a prominent $L$ window is observed to feature phonon hydrodynamics, we next investigate the spectral representation of lattice thermal conductivity ($\kappa_L$) in this $L$ window. In Fig. \ref{fig:kl_collective}, using the KCM approach, we present a spectral representation of $\kappa_L$, distinguished by its kinetic ($\kappa_{kinetic}$) and collective contributions ($\kappa_{collective}$), as a function of phonon frequency at $T$ = 10K for four different $L$. We choose $T$ = 10 K as a representative temperature to feature collective transport of phonons. The four different $L$ values have been chosen such that it covers a wide range that traverses from ballistic transport to the hydrodynamic regime at $T$ = 10 K. As we gradually increase the $L$ [from Figs. \ref{fig:kl_collective}(a) to \ref{fig:kl_collective}(d)), a gradual increment of the contributions coming from the collective transport is observed (shown via the red shaded regions inside the curve). The spectral $\kappa_L$ goes to zero before 2.87 THz, indicating the sole contribution of acoustic phonons in thermal transport at 10 K, as was realized earlier in Fig. \ref{fig:AM_OM_contrib_ballistic}(a).

To quantify the collective motion as a function of temperature for different $L$, we investigate the variation of characteristic non-local length ($l$) in GeTe at different temperatures and grain sizes. In a complete hydrodynamic description of thermal transport, the extension of the Guyer and Krumhansl equation \cite{Guyer1966_1} done in the KCM framework \cite{alvarez2018thermalbook}, namely, the hydrodynamic KCM equation, yields
\begin{equation}
    \tau \frac{\textit{d}\textbf{Q}}{\textit{dt}}+ \textbf{Q} = -\kappa \nabla T + l^{2}\left( \nabla^{2}\textbf{Q} + 2\nabla \nabla \cdot \textbf{Q} \right)
\end{equation}
where $\tau$ is the total phonon relaxation time, $\textbf{Q}$ is the heat flux, $\kappa$ is phonon thermal conductivity, and $l$ is the non-local length, that determines the non-local range in 
\onecolumngrid
\begin{widetext}
\vspace{-1cm}
\begin{figure}[H]
    \centering
    \includegraphics[width=1.0\textwidth]{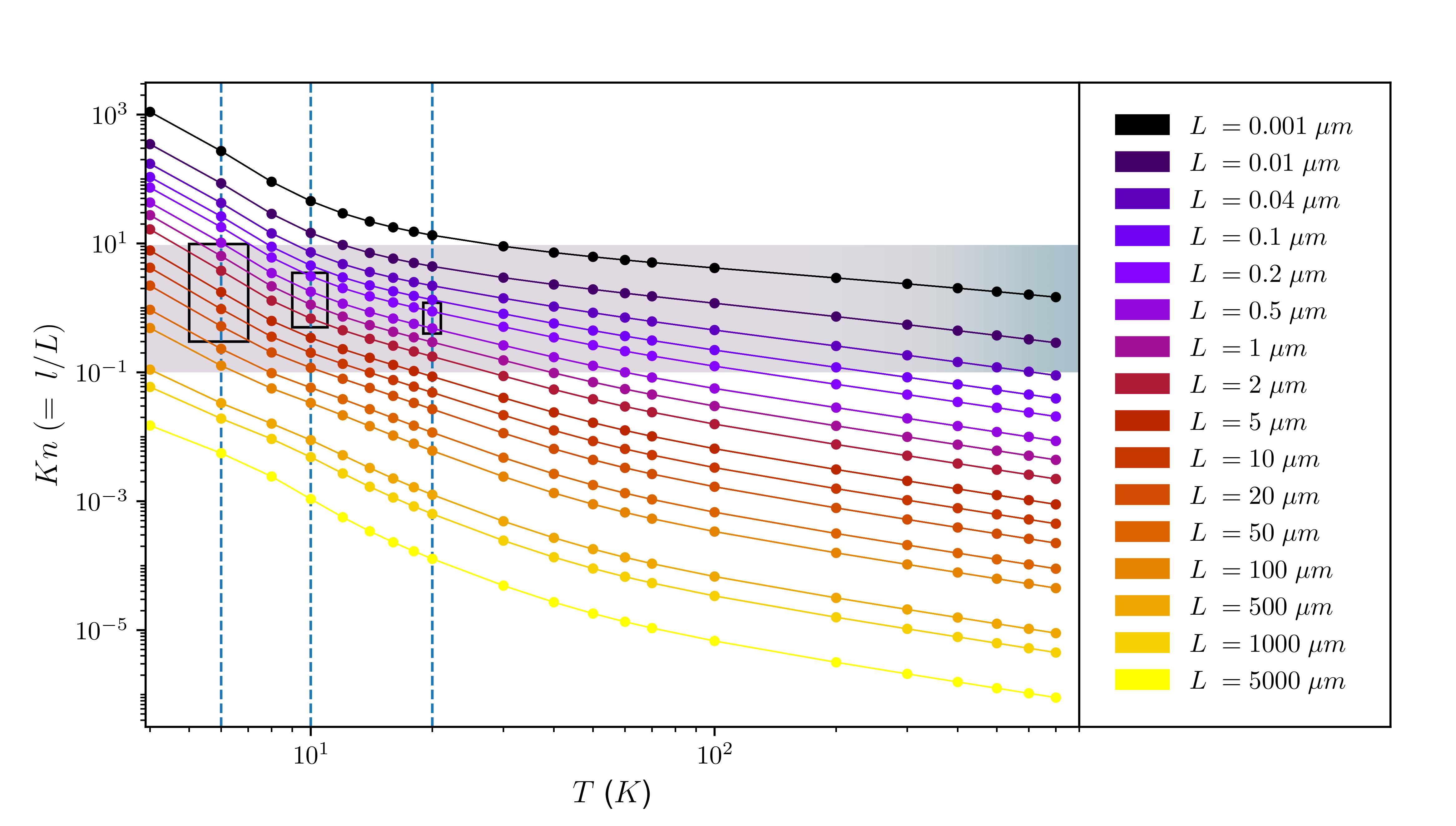}
    \caption{The variation of Knudsen number (Kn) with temperature for different $L$ values of crystalline GeTe. The shaded region satisfies 0.1 $\leq$ Kn $\leq$ 10 while the rectangular boxes define phonon hydrodynamic regimes calculated from average scattering rates. Blue dashed lines to guide the eye for $T$ = 6 K, 10 K and 20 K.}
    \label{fig:knudsen}
\end{figure}
\vspace{-0.5cm}
\end{widetext}

phonon transport. If we employ the steady state, strong geometric effects, and neglect the term $2\nabla \nabla \cdot \textbf{Q}$, then the equation can be visualized as analogous to Navier-Stokes equation with $l$ resembling heat viscosity. The Knudsen number (Kn) can be obtained from $Kn = l/L$ to study the collective motion quantitatively. Lower values of Kn define the recovery of Fourier's law whereas the hydrodynamic behavior becomes prominent when Kn gets higher values \cite{alvarez2018thermalbook, GUO20151}. Figure \ref{fig:knudsen} presents the variation of Kn as a function of temperature for different $L$. As temperature is lowered, a gradual increment of Kn is observed, concomitant with the gradual prominence of non-local behavior. Kn has earlier been described \cite{GUO20151, Bi2018} to indicate a phonon hydrodynamic regime when 0.1 $\leq$ Kn $\leq$ 10, bearing similarities with fluid hydrodynamics. We denote this region via a shaded region in Fig. \ref{fig:knudsen}. In Fig. \ref{fig:knudsen}, we also superpose the hydrodynamic $L$-window, identified using average scattering rates following Guyer and Krumhansl conditions for three representative temperatures: $T$ = 6 K, 10 K and 20 K. We observe that both definitions match well and the hydrodynamic $L$-window obtained by scattering rate analysis falls within the Kn range for hydrodynamics.

Knudsen number calculation also reveals the Ziman hydrodynamic regime for GeTe. Looking at the vertical dashed lines corresponding to $T$ = 6 K and $T$ = 10 K in Fig. \ref{fig:knudsen}, a small $L$-region is observed which does not fall into the rectangles, defined to indicate a hydrodynamic regime using scattering rate hierarchy. However, they fall inside the regime of 0.1 $<$ Kn $<$ 10, especially in the regime where Kn is close to 0.1. This corresponds to the Ziman hydrodynamic regime which corresponds to a regime where N scattering dominates but dissipates mostly by R scattering contrary to the Poiseuille hydrodynamic regime where N scattering dissipates mostly by the boundary scattering of the phonons. On the other hand, looking at $L$ values that lie inside 0.1 $<$ Kn $<$ 10 but with values close to 10, also sometimes do not lie inside the rectangular region (see the case of $L$ = 0.04 and 0.1 $\mu$m at $T$ = 10 K in Fig \ref{fig:knudsen}). Recalling Fig \ref{fig:scattering_rate}.(a), we observe that $L$ = 0.04 $\mu$m at $T$ = 10 K designates a scattering rate hierarchy, where $\langle \tau_{B}^{-1} \rangle_{ave} > \langle \tau_{N}^{-1} \rangle_{ave} > \langle \tau_{R}^{-1} \rangle_{ave}$, but $\langle \tau_{B}^{-1} \rangle_{ave}$ is not $\gg$ $\langle \tau_{ph-ph}^{-1} \rangle_{ave}$. Therefore, though it follows the prescribed hierarchy for hydrodynamics, the $L$ values do not enable a complete ballistic propagation and a competition between ballistic and diffusive phonons operates. This competition makes it difficult to distinguish sharp boundaries between different regimes. We will discuss more about this competition later.   
To characterize the repopulation of phonons in a different way, following Markov \textit{et al.} \cite{Bi2018}, we extract a length scale related to the propagation of heat wave before being dissipated, called the heat wave propagation length ($L_{hwpl}$), defined as a length at which the completely diffusive thermal conductivity decays 1/e times:
\begin{equation}
    \kappa_{L}(T, L)\mid_{L = L_{hwpl}} = \kappa_{L}(T, L > L_{diff})/e 
\end{equation}  
\begin{figure}[H]
    \centering
    \includegraphics[width=0.5\textwidth]{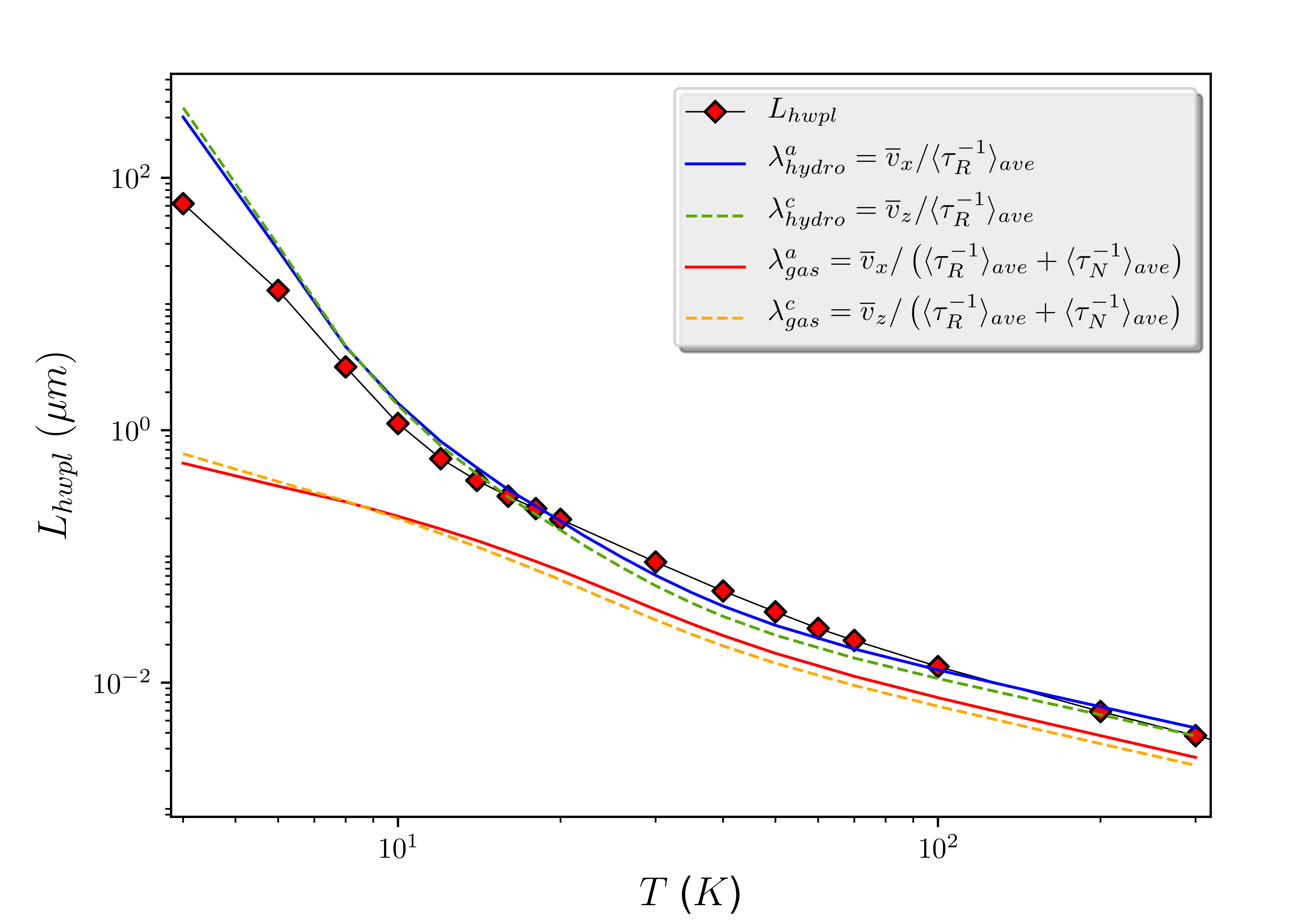}
    \caption{Heat wave propagation length ($L_{hwpl}$) as a function of temperature for crystalline GeTe. The temperature variation of phonon propagation lengths, correspond to the damping due to resistive scattering ($\lambda_{hydro}$) and both resistive and normal scattering ($\lambda_{gas}$) along $\textit{a}$ and hexagonal $\textit{c}$-axis are also presented.}
    \label{fig:hwpl}
\end{figure}

where $L_{diff}$ is the minimum length $L$, above which $\kappa_L$ reaches the thermodynamic limit, as mentioned earlier in Fig. \ref{fig:L_ball_L_diff}. $L_{hwpl}$ is connected to second sound, a typical characteristic for hydrodynamic heat transport phenomenon, which demonstrates the heat propagation as damped waves in a crystal \cite{Guyer1966_2, Cepellotti2015, umklapp_gang_2018} as a result of coherent collective motion of phonons due to the domination of N scattering. In this context, drift velocity of phonons ($\overline{v}$) and phonon propagation length ($\lambda_{ph}$) are defined as  
\begin{equation}
    \overline{v}_{j}^{2} = \frac{\sum_{\alpha} C_{\alpha}\mathbf{v}_{\alpha j}^{g}\cdot \mathbf{v}_{\alpha j}^{g}}{\sum_{\alpha}C_{\alpha}}
\end{equation}
and 
\begin{equation}
    \lambda_{ph} = \overline{v}/\langle \tau^{-1} \rangle _{ave}
\end{equation}
where, $C_{\alpha}$ is heat capacity of mode $\alpha$, $\mathbf{v}_{\alpha j}^{g}$ is phonon group velocity of mode $\alpha$ and $j$ can be either the component along the $a$ axis ($x$) or the hexagonal $c$ axis ($z$). Heat transfer of GeTe is anisotropic, as can be recalled from our earlier studies \cite{kanka,kanka2}, featuring different group velocities along the hexagonal $c$ axis and its perpendicular ($a$ axis) direction and therefore yields different drift velocities and different phonon propagation lengths along $x$ and $z$. Figure \ref{fig:hwpl} presents the variation of heat wave propagation length ($L_{hwpl}$) with temperature along with the variation of phonon propagation lengths along $x$ and $z$. Phonon propagation lengths are distinguished \cite{Bi2018} as $\lambda_{hydro}$ and $\lambda_{gas}$ via 

\begin{equation}
    \lambda_{hydro} = \overline{v}/\langle \tau_{R}^{-1} \rangle _{ave}
\end{equation}
\begin{equation}
    \lambda_{gas} = \overline{v}/\left(\langle \tau_{R}^{-1} \rangle _{ave}+\langle \tau_{N}^{-1} \rangle _{ave}\right)
\end{equation}

Figure \ref{fig:hwpl} shows the variation of heat wave propagation length ($L_{hwpl}$), superimposed with phonon propagation lengths with temperature along both $a$ and $c$ axis directions of GeTe. We observe that $L_{hwpl}$ follows well the trend of $\lambda_{hydro}$ as a function of temperature in the whole temperature range studied. $\lambda_{gas}$, the phonon propagation length corresponds to the uncorrelated phonon gas where both N and R scattering processes contribute to the damping of heat wave, on the other hand, seems to diverge from $L_{hwpl}$ as the temperature is lowered. This feature is an indication of gradual prominence of hydrodynamic behavior of phonons as the temperature is lowered. Similarly, the reasonable match between $L_{hwpl}$ and $\lambda_{hydro}$ predicts that heat wave propagation length is well captured by phonon flow with resistive damping caused by umklapp and isotope scattering. At very low temperature ($T$ = 4 K), a slight deviation is observed between $L_{hwpl}$ and $\lambda_{hydro}$ which can be attributed to the importance of boundary scattering as a significant damping process at very low temperature. 

Therefore, $L_{hwpl}$ can lead to the identification of the length scale at different temperatures at which phonon hydrodynamics can exist and therefore Poiseuille's flow and second sound phenomena can be observed. Interestingly, comparing $L_{hwpl}$ and characteristic size ($L$) of the sample, we can define Knudsen number in another approach as \cite{Bi2018} Kn = $L_{hwpl}$/$L$. The variation of Kn obtained using $L_{hwpl}$, is presented as a function of temperature in the Supplemental Material (Fig. S2). The variation of Kn with $T$ is found similar to our earlier evaluation of Kn using nonlocal length (Fig. \ref{fig:knudsen}).   

The blurry regions of transitions between ballistic, hydrodynamic, and diffusive transport are intriguing to understand the competition between different phonons with a wide range of mean free paths. Ideally, phonons with a wide spectrum of mean-free paths can be distinguished as either ballistic (MFP $>$ $L$) or diffusive (MFP $<$ $L$) phonons. However, the relative strength between ballistic and diffusive phonons are crucial to realize the competition between these two kind of phonons which eventually plays a decisive role in dictating the visible hydrodynamic phenomena. The phonon Knudsen minimum is such an indicator for the transition between ballistic and hydrodynamic phonon propagation regimes and had been used for several materials including graphene \cite{Li_graphene_2019}, graphite \cite{Ding2018}, SrTiO$_3$ \cite{Strontium2018}, black phosphorus \cite{Machida309} to detect phonon hydrodynamics. Figures \ref{fig:knudsen_minimum}.(a) and (b) present the the variation of normalized thermal conductivity ($\kappa_L^*$ = $\kappa_{L}/L$), a quantity that is similar to dimensionless $\kappa_L$, as a function of inverse Knudsen number, calculated using nonlocal length and heat wave propagation lengths respectively. Figure \ref{fig:knudsen_minimum}.(b) shows a wider range of 1/Kn as the Kn obtained using heat wave propagation length reaches higher values at low temperatures compared to that of the non-local length calculation from hydrodynamic KCM method. However, we observe almost similar trends of $\kappa_L^*$ with the variation of 1/Kn coming out of the two different approaches in obtaining the Knudsen number. At $T$ = 300 K, a steep linear decreasing trend is observed which is associated with the diffusive phonon scattering events as phonons behave as uncorrelated gas particles and resistive scattering is prominent and dominating at this temperature.
\onecolumngrid
\begin{widetext}
\vspace{-1cm}
\begin{figure}[H]
    \centering
    \includegraphics[width=1.0\textwidth]{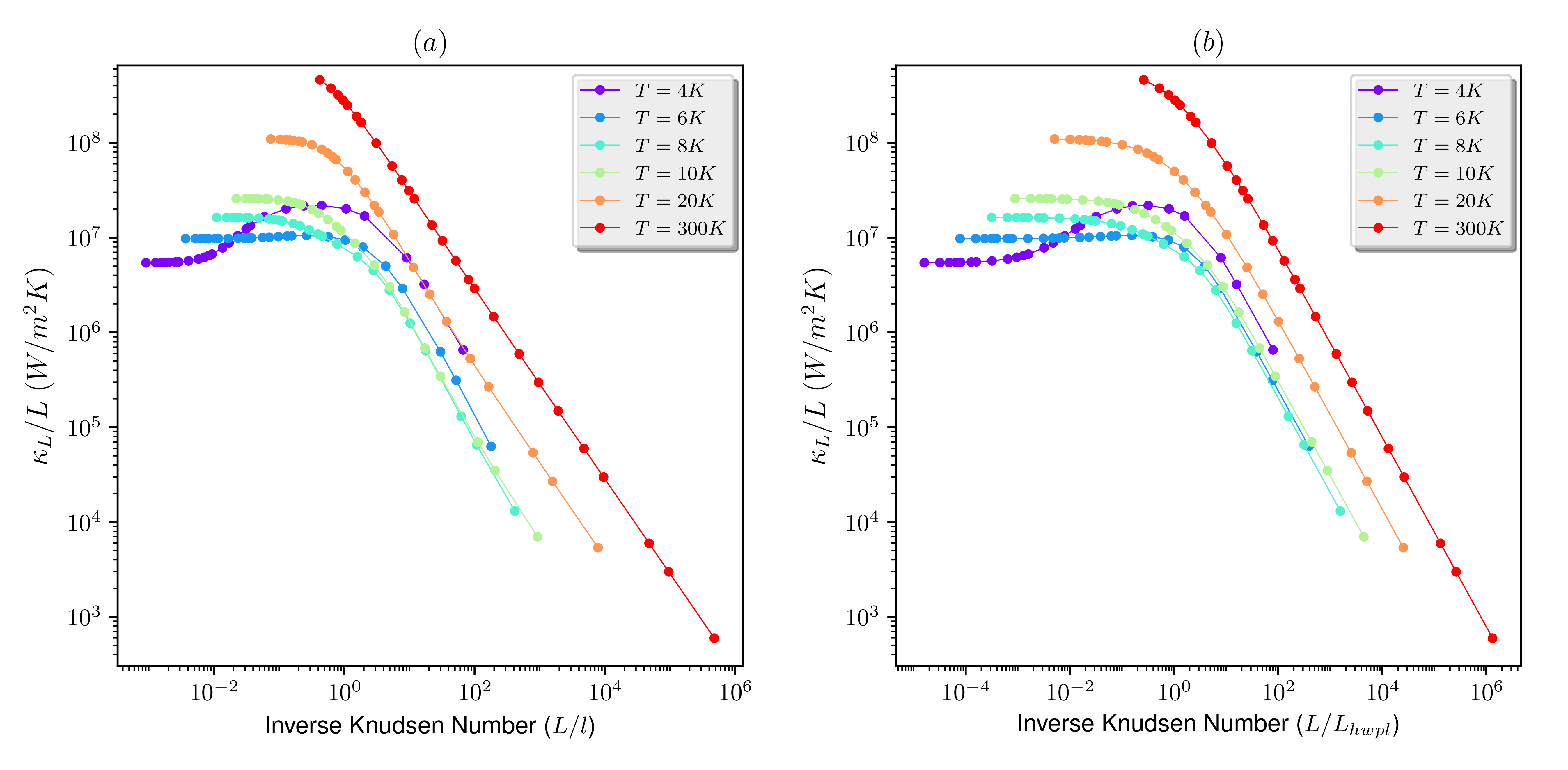}
    \caption{(a) The variation of normalized thermal conductivity ($\kappa_{L}/L$) as a function of inverse Knudsen number, calculated using characteristic non-local length for different temperatures. (b) The variation of normalized thermal conductivity ($\kappa_{L}/L$) as a function of inverse Knudsen number, calculated using heat wave propagation length for different temperatures.}
    \label{fig:knudsen_minimum}
\end{figure}
\vspace{-0.5cm}
\end{widetext}
    
\onecolumngrid
\begin{widetext}
\begin{figure}[H]
    \centering
    \includegraphics[width=1.0\textwidth]{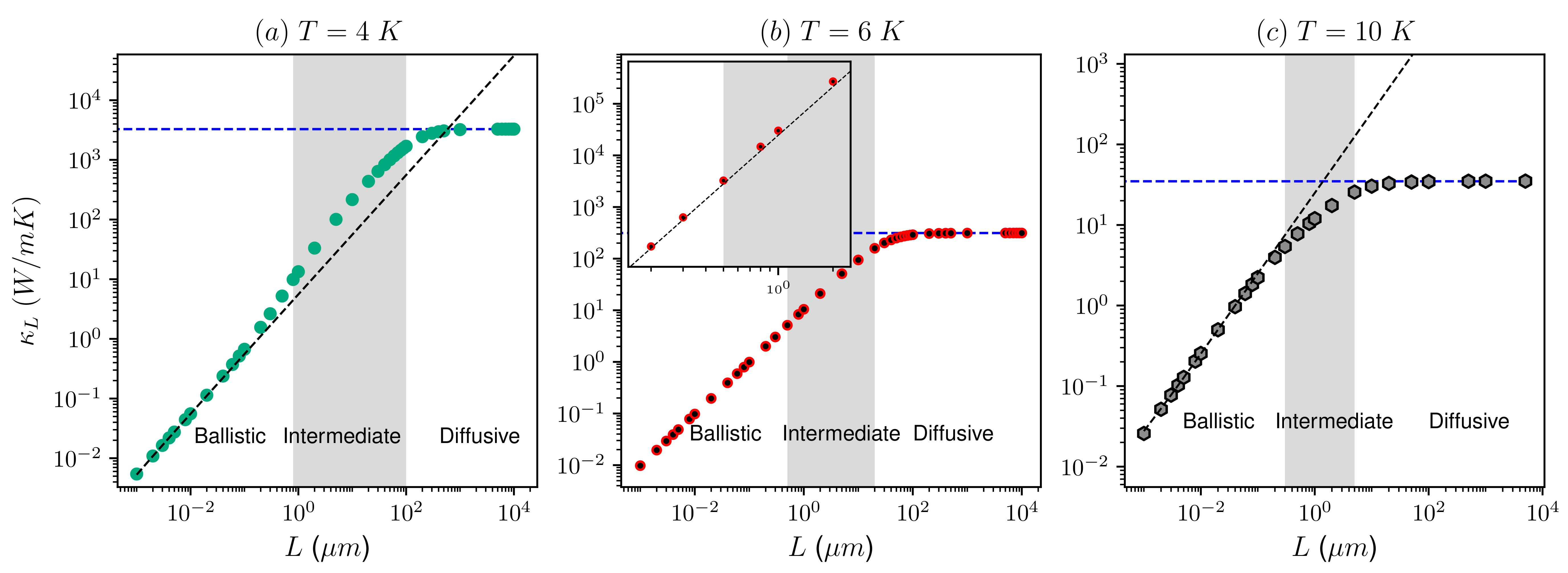}
    \caption{The variation of lattice thermal conductivity ($\kappa_L$) as a function of characteristic length ($L$) in log-log scale for different temperatures: (a) $T$ = 4 K, (b) $T$ = 6 K and (c) $T$ = 10 K. The inset of Fig \ref{fig:internediate}.(b) refers to the zoomed in view around linear to superlinear scaling at $T$ = 6 K. The intermediate regimes, located in between the ballistic and diffusive propagation regimes are shown via gray shades.}
    \label{fig:internediate}
\end{figure}
\vspace{-0.3cm}
\end{widetext}    
    
Starting from $T$ = 20 K, a gradual onset of a horizontal regime is visible before the linearly decreasing trend of $\kappa_L^*$  as the temperature is lowered. At $T$ = 4 K, surprisingly, a cusp-like trend, resembling that of a shallow minimum followed by a prominent maximum is observed before a linearly decreasing $\kappa_L^*$ at higher 1/Kn. The cusp-like shallow minimum at $T$ = 4 K indicates the phonon Knudsen minimum and predicts the presence of prominent transition from ballistic to hydrodynamic regime. Further, a prominent maximum in $\kappa_L^*$ has only been observed at $T$ = 4 K, which designates the strong presence of hydrodynamic phonon transport in GeTe. Similar observation can be found by Li \textit{et al.} \cite{Li_graphene_2019} for suspended graphene, where the increasing trend of $\kappa_L$, normalized by sample width, was attributed to the strong presence of hydrodynamic phonon transport.

The behavior of phonon Knudsen minimum of GeTe convinces us to understand the competition between ballistic and diffusive phonons in the quasi-ballistic regimes of phonon transport. We specifically turn our attention toward the reason behind the strong presence of hydrodynamics at $T$ = 4 K visible through Knudsen minimum in Fig \ref{fig:knudsen_minimum}. We recall that even $T$ = 6 K, $T$ = 8 K persist in phonon hydrodynamics, realized via the average scattering rate analysis and Knudsen number variation with temperature. To perceive the reason behind the difference between strong and weak phonon hydrodynamics, we investigate the scaling relation between $\kappa_L$ and $L$ in the intermediate regime of transport, where the transport is neither fully ballistic nor fully diffusive.       

Figure \ref{fig:internediate} describes the variation of $\kappa_L$ with $L$ at $T$ = 4 K, 6 K, and 10 K. Three phonon propagation regimes have been identified. At lower values of $L$, ballistic phonons dominate the transport and therefore a linear dependency of $\kappa_L$ on $L$ is observed. At very high $L$, the phonon transport is completely diffusive and a plateau-like regime is observed, denoting an independence of $\kappa_L$ over $L$. The intermediate regime where the phonon propagation shifts from complete ballistic to complete diffusive, plays a crucial role in determining the strong or weak presence of hydrodynamic propagation of phonons. Figure \ref{fig:internediate}.(c) indicates a sublinear variation in the intermediate regime at $T$ = 10 K. At $T$ = 6 K [Fig. \ref{fig:internediate}(b)], a minute superlinear behavior is observed while at $T$ = 4 K [Fig. \ref{fig:internediate}(a)], an enhanced superlinear behavior is perceived in the intermediate regime.

In the intermediate quasi-ballistic regime of phonon propagation, where both ballistic and diffusive phonons

\begin{figure}[H]
    \centering
    \includegraphics[width=0.5\textwidth]{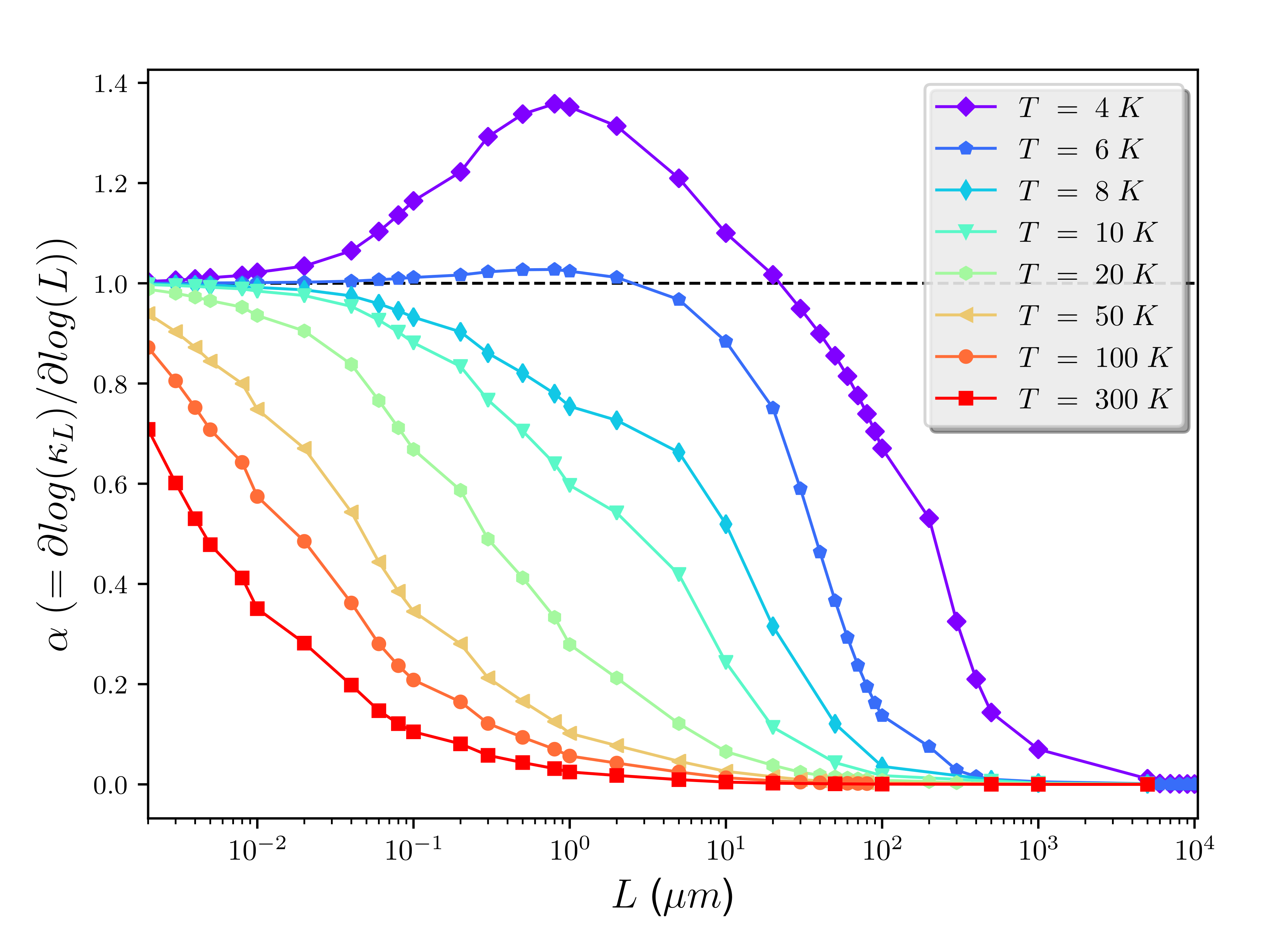}
    \caption{Variation of the scaling exponent $\alpha$ as a function of $L$ for different temperatures. The black dashed line denotes the $\alpha$ = 1 line.}
    \label{fig:alpha}
\end{figure}

operate and compete with each other, seems to be a marker to designate sample sizes ($L$) with strong hydrodynamic phonon transport characteristics. To further quantify the intermediate nonliearity (both sub and superlinearity), we evaluate and present the scaling exponent\cite{Ding2018} $\alpha$ = $\partial log (\kappa_L)$/$\partial log(L)$ as a function of $L$ for different temperatures in Fig. \ref{fig:alpha}. 

$\alpha$ = 0 indicates the size-independent behavior of $\kappa_L$ and therefore describes the completely diffusive phonon propagation regime. On the other hand, $\alpha$ = 1 reveals the linear size dependency and henceforth the complete ballistic phonon conduction regime. The superlinear dependence of $\kappa_L$ on $L$ in the intermediate regime can be captured by the the condition $\alpha$ $>$ 1. From Fig. \ref{fig:alpha} we observe that at low $L$, for low temperatures, $\alpha$ goes to 1. for higher temperatures, as expected almost no ballistic regime is observed with $\alpha$ $<$ 1. As we increase $L$, in the intermediate regime, a gradual departure from $\alpha$ = 1 is observed. For $T$ = 4 K and $T$ = 6 K, this departure leads to a regime with $\alpha$ $>$ 1, while for $T$ = 8K and 10 K this deviation leads to sublinear or $\alpha$ $<$ 1 scaling. At high $L$ values gradually all phonons become diffusive and $\alpha$ goes to zero.

There are several striking features to point out from Fig. \ref{fig:alpha}. First, prominent contribution of drifting phonons at 4 K leads to an enhanced superlinear scaling with $\alpha$ $>$ 1, representing the signature of phonon Poiseuille flow \cite{Ding2018} and therefore prominent phonon hydrodynamics which assists in featuring the Knudsen minimum seen in Fig. \ref{fig:knudsen_minimum}. Here we mention that even for $T$ = 4 K, the exponent $\alpha$ gradually starts from 1, reaches a maximum value around $L$ = 0.8 $\mu$m, and goes sublinear with $\alpha$ $<$ 1 thereafter before going to zero at very high $L$ values. Therefore, sublinear scaling is universal in the intermediate regime. For $T$ = 4 K, however, the sublinear scaling precedes a superlinear behavior displaying strong hydrodynamic feature. Second, a minute superlinear scaling, observed in Fig. \ref{fig:internediate}(b) inset for $T$ = 6 K, can be realized in a better way by observing the small $L$-window for which $\alpha$ $>$ 1 for $T$ = 6 K. At $T$ = 8K and 10 K, though sublinear scaling is observed in the intermediate regime, it decays to zero in different rates. After $L$ = 10 $\mu$m, the decay rate seems faster than that of the cases below $L$ = 10 $\mu$m.

We understand that although average scattering rate and Knudsen number variation with temperature indicates phonon hydrodynamics to present in GeTe for several temperature and characteristic size window, low-$\kappa_L$ material GeTe needs several factors to manifest a strong hydrodynamic response by phonons. In this context, superlinear scaling of $\kappa_L$ on $L$ plays a crucial role in the transition from complete ballistic to complete diffusive propagation regimes. To understand and investigate the reason behind superlinear and sublinear scaling at the intermediate quasi-ballistic regime of phonon transport, we calculate the ratio $\gamma$ as a function of $L$ for three temperatures: $T$ = 4 K, 6 K and 10 K.
We define $\gamma$ as 
\begin{equation}
    \gamma = \frac{\left\langle\tau_N^{-1}\right\rangle}{\left\langle\tau_R^{-1}\right\rangle + \left\langle\tau_B^{-1}\right\rangle}
\end{equation}
where $\left\langle\tau_N^{-1}\right\rangle$, $\left\langle\tau_R^{-1}\right\rangle$, and $\left\langle\tau_B^{-1}\right\rangle$ are average scattering 

\onecolumngrid
\begin{widetext}
\vspace{-1cm}
\begin{figure}[H]
    \centering
    \includegraphics[width=1.0\textwidth]{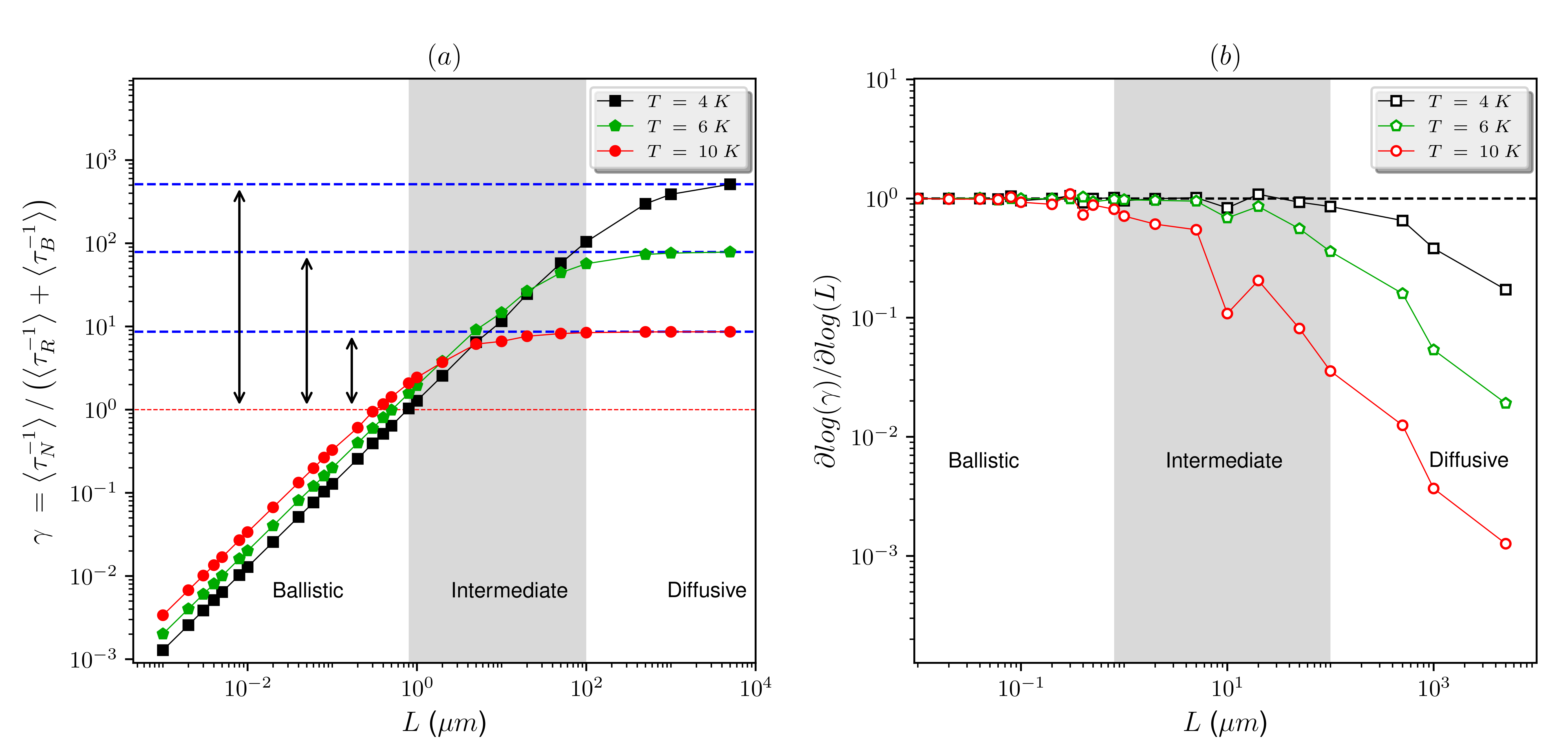}
    \caption{\label{fig:gamma} (a) $L$ dependence of $\gamma$ for three representative temperatures: $T$ = 4 K, 6 K and 10 K. The saturation values of $\gamma$ ($\gamma_{diff}$) in the plateau regimes attained at higher $L$ values for different temperatures are denoted via blue dashed lines. The red dotted line represents $\gamma$ = 1 and the differences between $\gamma$ = 1 and $\gamma_{diff}$ are shown via black double headed arrows. (b) The variation of $\partial log(\gamma)/\partial log(L)$ as a function of $L$ for three representative temperatures: $T$ = 4 K, 6 K and 10 K.}
\end{figure}
\vspace{-0.5cm}
\end{widetext}

rates for normal, resistive, and boundary scattering respectively. Figure \ref{fig:gamma}.(a) shows that $\gamma$ increases gradually and reaches a plateau as we increase $L$. In the ballistic phonon conduction regime, we observe $\gamma (T = 4 K)$ $<$ $\gamma (T = 6 K)$ $<$ $\gamma (T = 10 K)$. This is due to the effect of strong boundary scattering at low temperature and low $L$. However, in the regime of complete diffusive propagation of phonons a reverse trend is observed: $\gamma (T = 4 K)$ $>$ $\gamma (T = 6 K)$ $>$ $\gamma (T = 10 K)$ as in this regime, $\gamma$ is independent of size. We define these saturation values as $\gamma_{diff}$. Again, the crucial crossover is observed in the intermediate $L$-regime. We also mark the difference between $\gamma$ = 1 and $\gamma_{diff}$ in Fig. \ref{fig:gamma}(a) via double headed arrows. This difference characterizes the relative strength of normal scattering compared to the dissipative resistive scattering and therefore indicates the strength for persisting coherent phonon flow.

However, we tend to understand the reason behind the nonlinear behavior of $\kappa_L$ at the intermediate $L$-regime. For this purpose, we present the variation of the exponent of $\gamma$ by calculating $\partial log(\gamma)/\partial log(L)$ as a function of $L$ in Fig. \ref{fig:gamma}(b). We observe that the exponent for $T$ = 4 K is higher and for both $T$= 4 K and 6 K, it stays around 1 ($\gamma$ being linearly increasing with $L$) in the intermediate regime. However, for $T$ = 10 K, $\partial log(\gamma)/\partial log(L)$ drops up to several orders (at $L$ = 10 $\mu$m, it drops around 10 times) compared to the $T$ = 4 K and $T$ = 6 K cases. Therefore, the higher the exponent $\partial log(\gamma)/\partial log(L)$ in the intermediate regime and closer to 1, the higher the chances of the collective phonon flow due to strong normal scattering. This eventually can lead to the strong appearance of phonon hydrodynamics with superlinear $L$ dependence of $\kappa_L$ and  prominent Knudsen minimum apart from other signatures born out of the assessment of Knudsen number and average scattering rate as a function of temperature.

\section{Summary and conclusions}{\label{section:summary}}

We employ KCM in conjunction with first-principles density functional calculations to investigate the effect of characteristic size ($L$) on collective phonon transport in low-thermal conductivity material GeTe. We observe phonon hydrodynamics in crystalline GeTe and identify the competitive effects of both temperature and $L$ on the collective phonon transport. As a first step, we distinguish heat transport regimes correspond to ballistic and completely diffusive phonon transport. These regimes have been identified as a function of both temperature and $L$. In the ballistic regime, the frequency dependence of phonon propagation is understood. Temperature is found to dominate over $L$ in deciding the excitation of acoustic and optical phonons. Even for very small $L$ values, correspond to ballistic transport regime, we observe a small contribution coming from optical modes of GeTe if temperature is raised to 30 K. However, at low temperature ($T$ = 10 K), only acoustic modes excite to enable ballistic propagation. The variation of mean free paths as a function of frequencies also represents this dependence. At low temperatures, increasing $L$ gradually weakens ballistic conduction. On the other hand, for the same $L$ value, increasing temperature also gradually weakens the ballistic conduction. 

After looking at ballistic and diffusive phonon conduction regimes, we turn our attention toward the intriguing intermediate $L$-regime where both ballistic and diffusive phonons are present. The average scattering rates seem to follow the Guyer and Krumhansl hierarchy at low temperatures, indicating the presence of phonon hydrodynamics at certain temperatures and $L$ window. KCM method allows us to distinguish the variation of collective contribution as functions of both temperature and $L$. Therefore, the phonon hydrodynamic regimes in terms of both temperature and $L$ have been realized using non-local length and Knudsen number (Kn) evaluation which draws a parallel between fluid hydrodynamics and the collective flow of phonons. The hydrodynamic regimes identified using scattering rates are found to satisfy the condition 0.1 $<$ Kn $<$ 10, which is the condition for hydrodynamic flow in terms of Kn. Further, exploiting the variation of lattice thermal conductivity as a function of $L$, a heat wave propagation length has been extracted for different temperatures. Comparing this characteristic length scale with phonon propagation lengths reveals that the heat wave propagation length is well captured by phonon propagation with only resistive damping. the Knudsen number can also be associated with this length scale which shows almost similar behavior as that of the Kn obtained using a nonlocal length. For both of these definitions of Kn, the variation of normalized thermal conductivity ($\kappa_L^{*}$ = $\kappa_{L}/L$) with 1/Kn shows a Knudsen minimum like feature only at very low temperature ($T$ = 4 K). Though Kn can capture the hydrodynamic regimes well in terms of both temperature and $L$, some of the prominent features of phonon hydrodynamics, like Knudsen minimum, can be weakly present or may be absent in low-thermal conductivity materials. We have found that the intermediate $L$-regime and the scaling of thermal conductivity with $L$ in this regime works as a marker to determine the existence of the Knudsen minimumlike prominent hydrodynamic feature. A superlinear scaling in this intermediate $L$-regime seems to assist a Knudsen minimum and therefore prominent phonon hydrodynamics. On the other hand, sublinear scaling does not lead to a Knudsen like minimum, featuring weak phonon hydrodynamics at those temperatures. A ratio of average normal and resistive scattering rates have been found to control the strength and prominent visibility of the collective phonon transport in GeTe. 

In summary, this paper reveals crucial details about how and when the prominent signatures of phonon hydrodynamics can be observed in low-thermal conductivity materials. In this context, it demonstrates and systematically analyzes the consequences of the competitive effects between temperature and characteristic size on phonon hydrodynamics in GeTe. The outcome of this study can lead to a better understanding of the generic behavior and consequences of the phonon hydrodynamics and its controlling parameters in any other low-thermal conductivity materials. The accurate description of phonon hydrodynamics in low-$\kappa$ materials can also lead to better theoretical predictions of experimentally observed thermal conductivity at low temperatures for these materials.

\begin{acknowledgments}

This project has received funding from the European Union’s Horizon 2020 research and innovation program under Grant Agreement No. 824957 (“BeforeHand:” Boosting Performance of Phase Change Devices by Hetero- and Nanostructure Material Design).
\end{acknowledgments}

\onecolumngrid

\begin{widetext}
\section{Supplementary Material}
\end{widetext}

\subsection{Collective and Kinetic mean free path}

Figure \ref{fig:mfp1} presents the variation of effective mean free paths of GeTe as a function of phonon frequency for two different $L$ values at $T$ = 10 K. In KCM approach, mean free paths can be distinguished as kinetic and collective mean free paths. As described in the main text, the kinetic mean free paths ($l_{k} (\omega)$) are different for different modes but the collective mean free paths ($l_{c} (T)$) are same for all modes and it is only a function of temperature. The effective mean free path has been realized using $l_{eff} (\omega)$ = (1-$\Sigma$)$l_{k}(\omega)$+$\Sigma$ $l_{c}$. In Fig \ref{fig:mfp1}.(a), we observe that at $T$ = 10 K, the dominating contribution comes from the kinetic mean free path for $L$ = 0.003 $\mu$m. We recall from the main text that this $L$ corresponds to ballistic phonon conduction regime for GeTe at 10 K. However, $L$ = 0.8 $\mu$m satisfies the Guyer and Krumhansl condition for phonon hydrodynamics and consistently the collective mean free path is seen to dominate over the kinetic mean free path (Fig \ref{fig:mfp1}.(b)). 

\onecolumngrid
\begin{widetext}
\makeatletter 
\renewcommand{\thefigure}{S\@arabic\c@figure}
\makeatother
\setcounter{figure}{0}
\begin{figure}[H]
    \centering
    \includegraphics[width=1.0\textwidth]{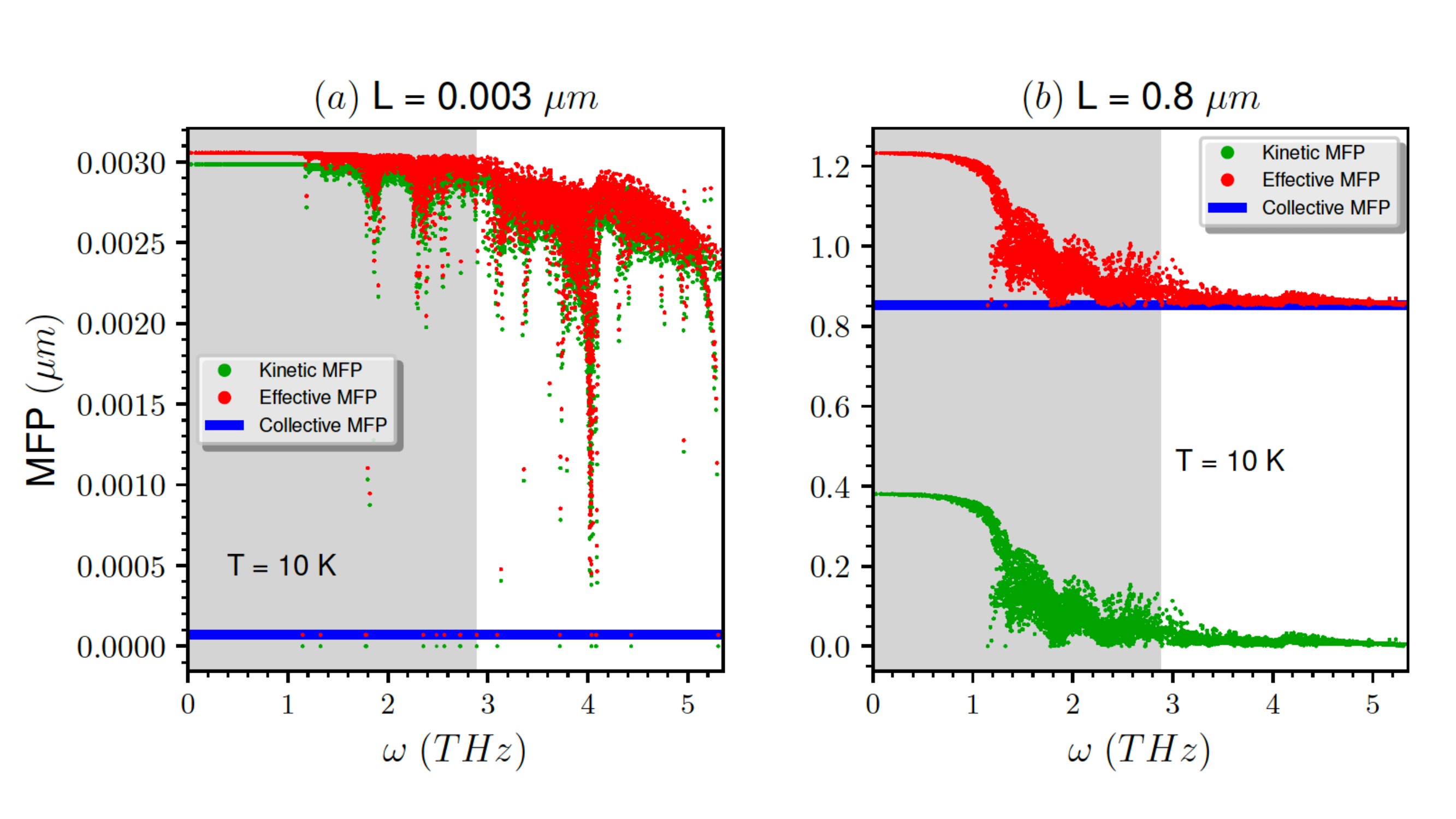}
    \caption{Effective mean free path (MFP) of crystalline GeTe, along with collective and kinetic contributions are presented as a function of frequencies for two different $L$ at $T$ = 10 K: (a) $L$ = 0.003 $\mu$m and (b) $L$ = 0.8 $\mu$m. Gray shaded regions denote acoustic mode frequency regime for GeTe.}
    \label{fig:mfp1}
\end{figure}

\end{widetext}

\subsection{Variation of Knudsen number with temperature for different $L$ using heat wave propagation length}

Figure \ref{fig:knudsen2} presents the variation of Knudsen number (Kn) as a function of temperature for different $L$ values. Kn has been realized via the heat wave propagation length ($L_{hwpl}$) as $L_{hwpl}$/$L$, where $L_{hwpl}$ is obtained as a characteristic length at which the lattice thermal conductivity in the completely diffusive limit correspond to bulk sample reduces to 1/e times. The hydrodynamic regime follows 0.1 $<$ Kn $<$ 10 and therefore has been marked with the shaded region. The hydrodynamic $L$-regimes obtained using average scattering rates are also shown via rectangular boxes for $T$ = 6 K, 10 K and 20 K. Except the ballistic hydrodynamic boundary regimes for $T$ = 6 K, the regimes evaluated by Kn and average scattering rates are found to be consistent. The transition between ballistic and hydrodynamic regimes are often found to be blurry and without sharp demarcation in low-thermal conductivity materials. This has been discussed in the main text.  

\onecolumngrid
\begin{widetext}
\makeatletter 
\renewcommand{\thefigure}{S\@arabic\c@figure}
\makeatother
\setcounter{figure}{1}
\begin{figure}[H]
    \centering
    \includegraphics[width=1.0\textwidth]{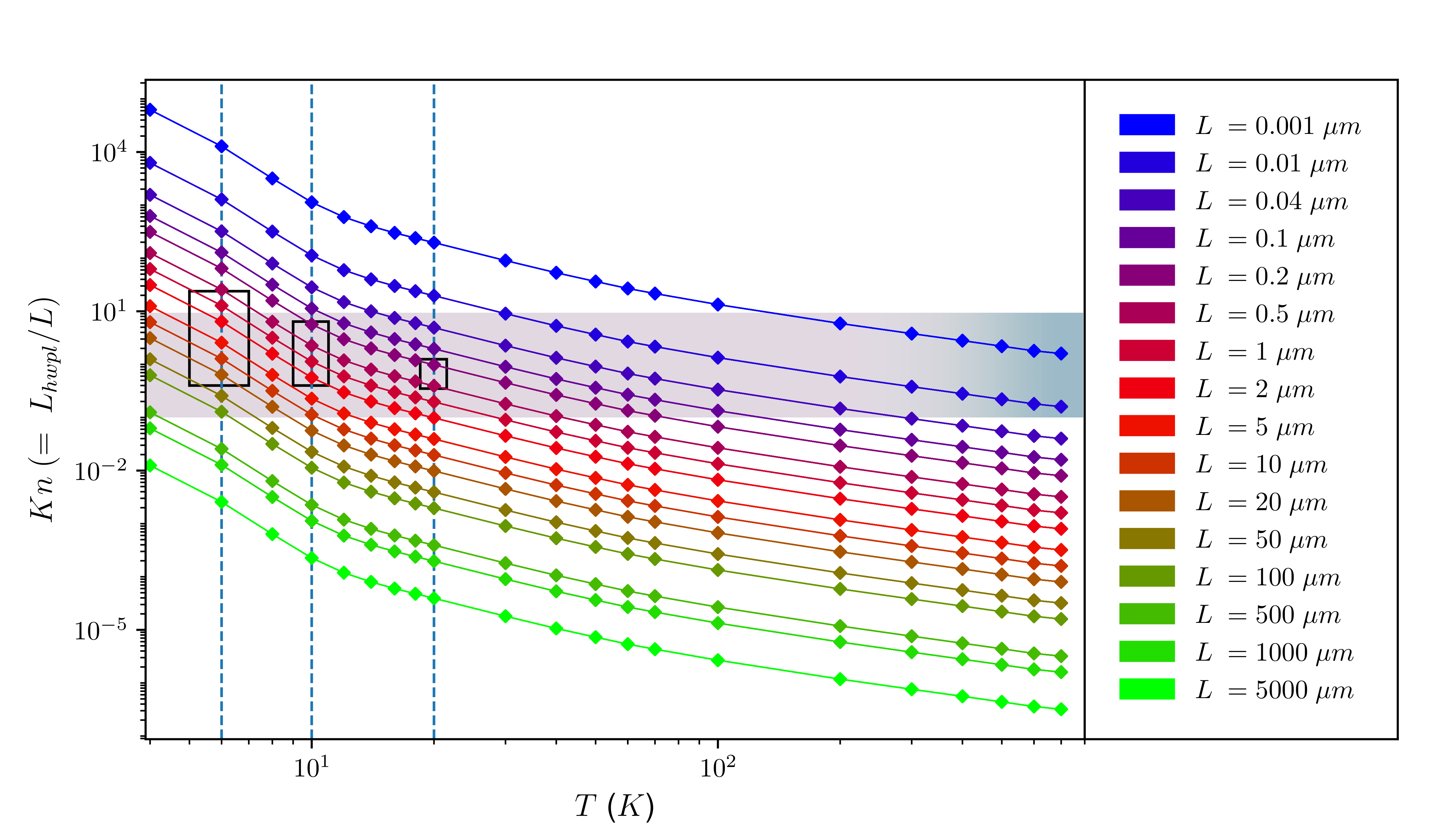}
    \caption{The variation of Knudsen number (Kn) with temperature for different $L$ values of crystalline GeTe. The shaded region satisfies 0.1 $\leq$ Kn $\leq$ 10 while the rectangular boxes define phonon hydrodynamic regimes calculated from average scattering rates. Blue dashed lines to guide the eye for $T$ = 6 K, 10 K and 20 K.}
    \label{fig:knudsen2}
\end{figure}
\end{widetext}

\nocite{*}

\bibliography{apssamp}


\end{document}